\documentclass[pra, 10pt, letterpaper, twocolumn, aps]{revtex4-1}

\usepackage{graphicx,amsmath,amssymb}
\usepackage{color}
\usepackage[normalem]{ulem}
\usepackage{epstopdf}
\usepackage{lmodern}
\usepackage[utf8]{inputenc}
\usepackage{lipsum}
\usepackage[colorlinks=true,citecolor=blue,linkcolor=blue,urlcolor=blue]{hyperref}

\def\ket#1{\left|#1\right\rangle}
\def\bra#1{\left\langle#1\right|}
\def\braket#1{\left\langle#1\right\rangle}

\newcommand{\defeq}{\mathrel{\vcenter{\baselineskip0.5ex \lineskiplimit0pt
                     \hbox{\scriptsize.}\hbox{\scriptsize.}}}%
                     =}

\usepackage{tikz}
\usetikzlibrary{patterns}
\usetikzlibrary{shapes.geometric}
\usetikzlibrary{positioning,arrows}
\usetikzlibrary{decorations.pathmorphing}
\usetikzlibrary{decorations.markings}

\begin{document}
\title{Phase magnification by two-axis countertwisting for \\ detection-noise robust interferometry}

\author{Fabian~Anders,$^1$ Luca~Pezzè,$^2$ Augusto~Smerzi,$^2$ and Carsten~Klempt$^1$}
\affiliation{
$^1$ Institut f\"ur Quantenoptik, Leibniz Universit\"at Hannover,  Welfengarten~1, D-30167~Hannover, Germany \\
$^2$ QSTAR, INO-CNR and LENS, Largo Enrico Fermi 2, I-50125, Firenze, Italy
}

\begin{abstract}
Entanglement-enhanced atom interferometry has the potential of surpassing the standard quantum limit and eventually reaching the ultimate Heisenberg bound.
The experimental progress is, however, hindered by various technical noise sources, 
including the noise in the detection of the output quantum state.
The influence of detection noise can be largely overcome 
by exploiting echo schemes, where the entanglement-generating interaction is repeated after the interferometer sequence.
Here, we propose an echo protocol that uses two-axis countertwisting as the main nonlinear interaction.
We demonstrate that the scheme is robust to detection noise and 
its performance is superior compared to the already demonstrated one-axis twisting echo scheme.
In particular, the sensitivity maintains the Heisenberg scaling in the limit of a large particle number.
Finally, we show that the protocol can be implemented with spinor Bose-Einstein condensates. 
Our results thus outline a realistic approach to mitigate the detection-noise in quantum-enhanced interferometry.
\end{abstract}

\maketitle

\section{Introduction}
The fast progress in the field of atom interferometry is characterized by improving precision and accuracy, 
a transition to both large-scale and compact devices, and an increasing number of metrology and sensing 
applications \cite{Tino2014,CroninRMP2009, KitchingIEEESensors2011}. 
State-of-the-art atom interferometers are linear two-mode devices that
employ uncorrelated (or, at most, classically-correlated) particles. 
Their phase estimation uncertainty is bounded by the standard quantum limit (SQL), 
$\Delta \theta_{\rm SQL} = 1/\sqrt{\nu N}$, where
$\nu$ is the number of repeated measurements and $N$ the number of particles in each shot.
Interferometers using ensembles of entangled atoms
can surpass the SQL, up to the ultimate Heisenberg limit (HL), $\Delta \theta_{\rm HL} =1/\sqrt{\nu} N$~\cite{Toth2014, Pezz`e2016} .
Factual implementations of entanglement-enhanced interferometer schemes~\cite{Pezz`e2016} 
have reached the HL with only few particles ($N \lesssim 10$ trapped ions).
Large gain in phase sensitivity (up to a factor 10 over the SQL~\cite{Hosten2016, Cox2016}) has been demonstrated with 
cold atoms and Bose-Einstein condensates employing $N=10^2$-$10^6$ atoms~\cite{Pezz`e2016} . 
Yet, these experiments have neither reached the HL nor the Heisenberg scaling 
$\Delta \theta \sim 1/N$ of phase uncertainty with the atom number.
This is caused by difficulties in creating metrologically-useful entangled states of a large number of atoms as well as
unavoidable technical noise, in many cases dominated by the noise in the final detection of the interferometer output state.
In particular, it has been shown that---in typical atom detection scenarios---the finite measurement efficiencies impose a bound to the achievable phase sensitivity that scales as $1/\sqrt{N}$
for sufficiently large atom numbers, at best with a constant improvement factor over the SQL~\cite{Escher2011, Demkowicz2012}.

Interestingly, the requirement of high detection efficiencies can be relaxed when using active detection
schemes, where a nonlinear interaction between the probe particles is applied after the phase imprinting
(also indicated as interaction-based readout). 
A signal amplification based on a nonlinear interaction has been first proposed in a seminal publication by Caves~\cite{Caves1982},
where the displacement of a coherent state is amplified by degenerate parametric amplification.
Along this line, an interferometric setup
was proposed by Yurke \textit{et al.}~\cite{Yurke1986}---called a SU(1,1) interferometer---where a nonlinear interaction (such as optical parametric down-conversion) generates correlated pairs of particles in two side modes from a pump mode.
After a phase shift acquired by the side modes relative to the pump, the SU(1,1) interferometer is closed by an inverted down-conversion.
The phase shift can be inferred from the residual population of the side modes.
SU(1,1) interferometers can reach a sensitivity at the HL with respect to the number of particles in the side modes
after the first down-conversion~\cite{Yurke1986} and a Heisenberg scaling of sensitivity with respect to the total number of particles in the probe state~\cite{Gabbrielli2015}.
Recently, it has been shown that a modified version (called ``pumped-up'' SU(1,1) interferometer~\cite{Szigeti2017})
where the three modes are further linearly coupled before and after phase imprinting, 
can reach a sub-SQL sensitivity with respect to the total number of particles~\cite{Szigeti2017}.
The robustness of SU(1,1) interferometers with respect to detection noise has been emphasized \cite{Marino2012, Ou2012, Gabbrielli2015, Szigeti2017}.
SU(1,1) interferometry with Bose-Einstein condensates~\cite{Linnemann2016}, photons~\cite{Jing2011, Hudelist2014},
and hybrid atom-light systems \cite{ChenPRL2015} has been recently demonstrated experimentally.

Interaction-based readout can also be used in the more standard two mode linear SU(2) interferometers
\cite{Davis2016, Froewis2016, Nolan2017, Hosten2016a} as
successfully demonstrated experimentally with trapped ions prepared in a GHZ state 
\cite{Leibfried2005} and with cold atoms prepared in a
spin-squeezed state in an optical cavity~\cite{Hosten2016}.
Up to now, both the proposals \cite{Davis2016, Froewis2016, Nolan2017} 
and the implementations \cite{Leibfried2005, Hosten2016} in SU(2) interferometers rely on the one-axis twisting (OAT) interaction \cite{kitagawa1993}.

\tikzstyle{arrow} = [thick,->,>=stealth]
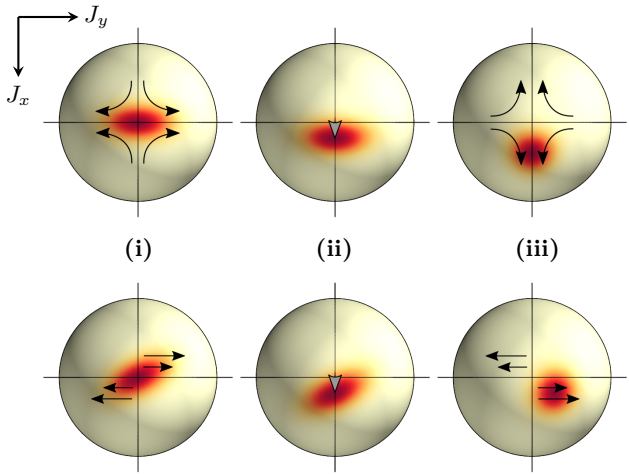
\begin{figure}[t!]
	\centering
  	\begin{tikzpicture}
	\node[inner sep=0pt] (a) at (0,1.7)
	    {\includegraphics[width=.46\textwidth]{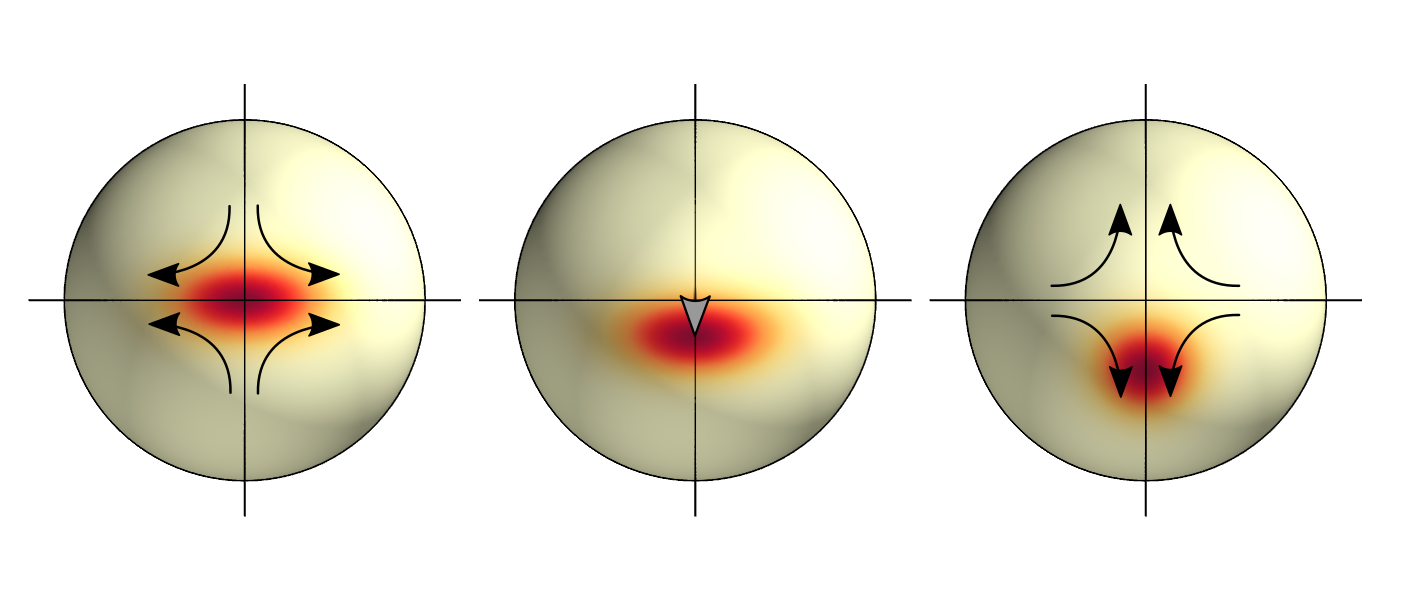}};
	\node[inner sep=0pt] (b) at (0,-1.7)
	    {\includegraphics[width=0.46\textwidth]{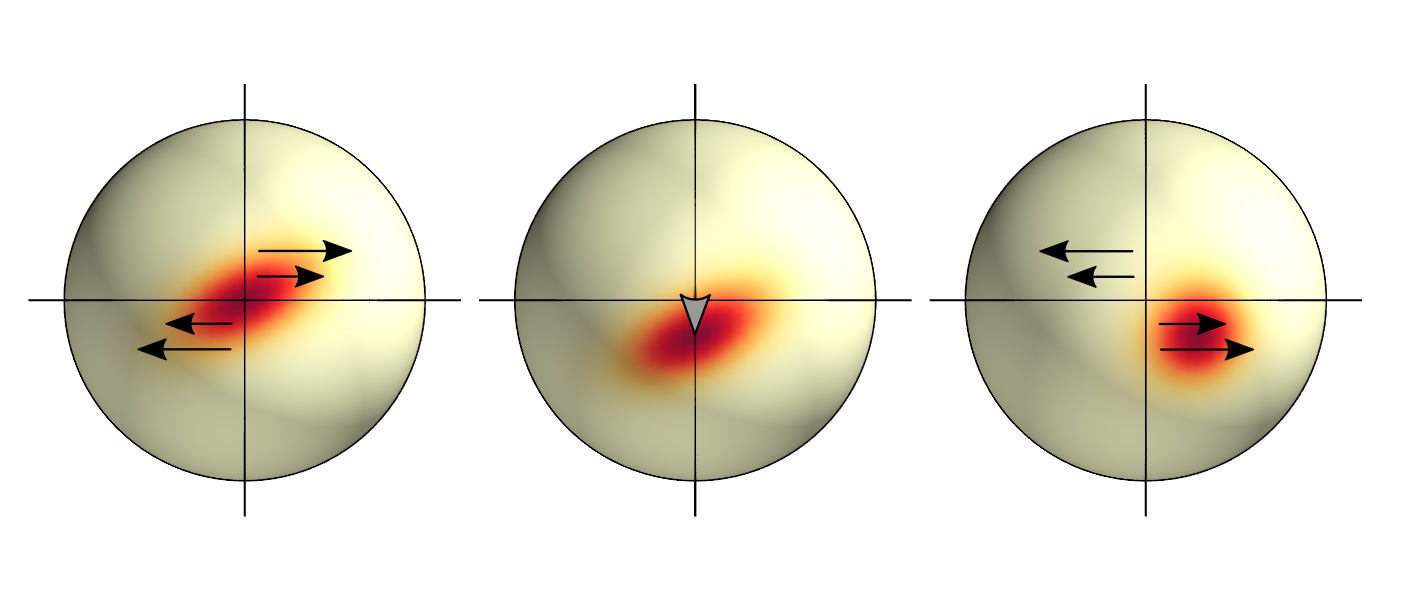}};
	\node[inner sep=0pt] (1) at (-2.7,0.0) {\small \textbf{(i)}};
	\node[inner sep=0pt] (2) at (-0.1,0.) {\small \textbf{(ii)}};
	\node[inner sep=0pt] (2) at (2.6,0.0) {\small \textbf{(iii)}};
	
	\draw [arrow] (-4.3,3.1) -- node[pos=1.3] {$J_y$} (-3.5,3.1);
	\draw [arrow] (-4.3,3.1) -- node[pos=1.3] {$J_x$} (-4.3, 2.3);	
	\end{tikzpicture}
	\caption[Schematic visualization of twisting-echo schemes.]{Schematic visualization (Husimi distributions)
	of the TACT echo (top row) and OAT echo (bottom row) protocols.
Black arrows visualize the action of the respective nonlinear Hamiltonian and correspond to classical trajectories.
(i) Preparation of a squeezed state (with $-6\, \mathrm{dB}$ squeezing) starting from a CSS with $N=100$ atoms.
(ii) Interferometric phase imprint (rotation around the $y$-axis) of $\theta = 0.2$.
(iii) Inverse nonlinear dynamics leads to anti-squeezing and phase magnification. 
For the TACT echo the magnification is aligned with the phase imprint direction, whereas for the OAT echo the magnification happens in the perpendicular direction.
}
	\label{fig1}
\end{figure}

In this manuscript, we propose an echo protocol based on two-axis 
countertwisting (TACT)~\cite{kitagawa1993, Andre2002, Liu2011, Yukawa2014, Kajtoch2015, Zhong2017}
that realizes an interaction-based readout and outperforms OAT proposals in both, the spin-squeezing and the highly over-squeezed regime.
The TACT echo benefits from a perfect alignment of squeezing and phase shift, leading to an 
ideal signal-to-noise ratio (SNR), a higher optimal phase sensitivity, an improved noise robustness, and a broader versatility.
For small magnification, the TACT echo realizes the simple single-mode linear amplifier described by Caves~\cite{Caves1982}.
The optimal achievable phase sensitivity of the TACT echo is only a factor $20\%$ below the HL and maintains a Heisenberg scaling even in the presence 
of number-dependent detection noise.
We present a realization of a robust TACT echo with spin dynamics in atomic Bose-Einstein condensates, 
where the TACT echo becomes similar to the pumped-up SU(1,1) protocol of Ref.~\cite{Szigeti2017}.
In particular, we extend the existing analysis showing that the pumped-up SU(1,1) protocol 
in the over-squeezed regime can reach a Heisenberg scaling of phase sensitivity with respect to the 
total number of particles, with large robustness to detection noise.

\section{TACT echo protocol}
We consider an ensemble of atoms occupying only two modes $\ket{a}$ and $\ket{b}$.
We assume that only states symmetric under particles exchange are available and introduce the
collective spin operator
\begin{align}\label{equ:pseudo-spin}
\begin{pmatrix}J_x \\ J_y \\ J_z \end{pmatrix}
&=
\frac{1}{2}\begin{pmatrix}a^\dagger b + b^\dagger a \\-i(a^\dagger b - b^\dagger a) \\ a^\dagger a - b^\dagger b \end{pmatrix}.
\end{align}
Here, $a^\dagger,b^\dagger$ ($a,b$ ) are the creation (annihilation) operators for the two modes. 
We also introduce $J_\pm = J_x \pm i J_y$.
The TACT  interaction is described by the Hamiltonian~\cite{kitagawa1993} 
\begin{align}
\label{equ:H_TACT}
H_{\rm TACT} = -\frac{\chi}{2i} \left( J_+^2 - J_-^2 \right)
= -\chi  \left( J_x J_y + J_y J_x \right),
\end{align}
and should be compared to the more familiar OAT model (here along the $x$ direction)~\cite{kitagawa1993}
 \begin{align}
\label{equ:H_OAT}
H_{\rm OAT}  = \chi J_x^2.
\end{align}
The echo scheme consists of the following transformation:
\begin{align}\label{equ:tact_echo_out}
	\ket{\psi}_\mathrm{out} = U^{-r} e^{-i\theta J_y } U \ket{\psi}_{\rm in},
\end{align}
where $\ket{\psi}_{\rm in}$ is an initial state [in the following we consider a coherent spin state (CSS) polarized along $J_z$] and $U = e^{-i t H}$, with $H$ given by Eq.~(\ref{equ:H_TACT}) for the TACT scheme, and by Eq.~(\ref{equ:H_OAT}) for the OAT scheme.
After phase encoding, the transformation $U^{-r}$
inverts the dynamics, with the factor $r$ allowing for a variable echo interaction.
The sequence of transformations Eq. (\ref{equ:tact_echo_out}) can be visualized on the Bloch sphere as shown in 
Fig.~\ref{fig1} for the TACT (top row) and for the OAT (bottom row):
(i) A spin-squeezed state is dynamically generated by $e^{-it H_{\rm TACT}}$ or $e^{-i tH_{\rm OAT}}$.
It should be noticed that while the Hamiltonian $H_{\rm TACT}$ creates a spin-squeezed state with reduced uncertainty along $J_x$,
$H_{\rm OAT}$ generates spin squeezing in the $x$-$y$ plane at an angle that depends on time~\cite{kitagawa1993}.
(ii) We apply an interferometric transformation described by the unitary $e^{-i\theta J_y}$, where $\theta$ is a phase shift.
Notice that, to be maximally sensitive to the transformation $e^{-i\theta J_y}$, the 
state generated by OAT must be first rotated in the Bloch sphere so to align the squeezed ellipse along the $y$ axis.
However, this alignment is generally not beneficial for the successive application of the OAT echo.
The state created by the TACT protocol is already optimally aligned for a rotation around $J_y$ and no additional transformation is required.
(iii) The echo scheme now comprises a characteristic second interaction $e^{+i t H_{\rm TACT}}$ or $e^{+i t H_{\rm OAT}}$ (here, $r=1$)
in the two cases, respectively, which is inverse to the first one.
In both cases, the echo interaction undoes the squeezing, returning (in the case $\theta \approx 0$) the state toward a CSS.
In the TACT case, the relevant uncertainty along $J_x$ is hereby increased, 
and the imprinted phase is magnified to a more substantial displacement from the north pole.
This second interaction maintains the signal-to-noise ratio because both the signal (the phase) and the noise (the spin uncertainty) are simultaneously amplified.
Compared to the OAT echo~\cite{Davis2016} (see Fig.~\ref{fig1}, bottom row), the TACT echo features an 
alignment of phase shift and squeezing direction, yielding an optimal and more clean phase magnification.
The comparison between OAT and TACT scheme will be discussed in more details below.
(iv, not visualized) The amplified phase shift is read out by a final detection of the output state.
For this purpose, the state is rotated by $\pi/2$ toward the equator of the Bloch sphere (around $J_y$ for the TACT echo and around $J_x$ for the OAT echo).
Thereby, the detection operates at mid-fringe position in terms of a final measurement of $J_z$, i.e., counting the atoms in the two modes.

\begin{figure}[t!]
  \centering
  	\begin{tikzpicture}
	\node[inner sep=0pt] (a) at (0,1.8)
	    {\includegraphics[width=.46\textwidth]{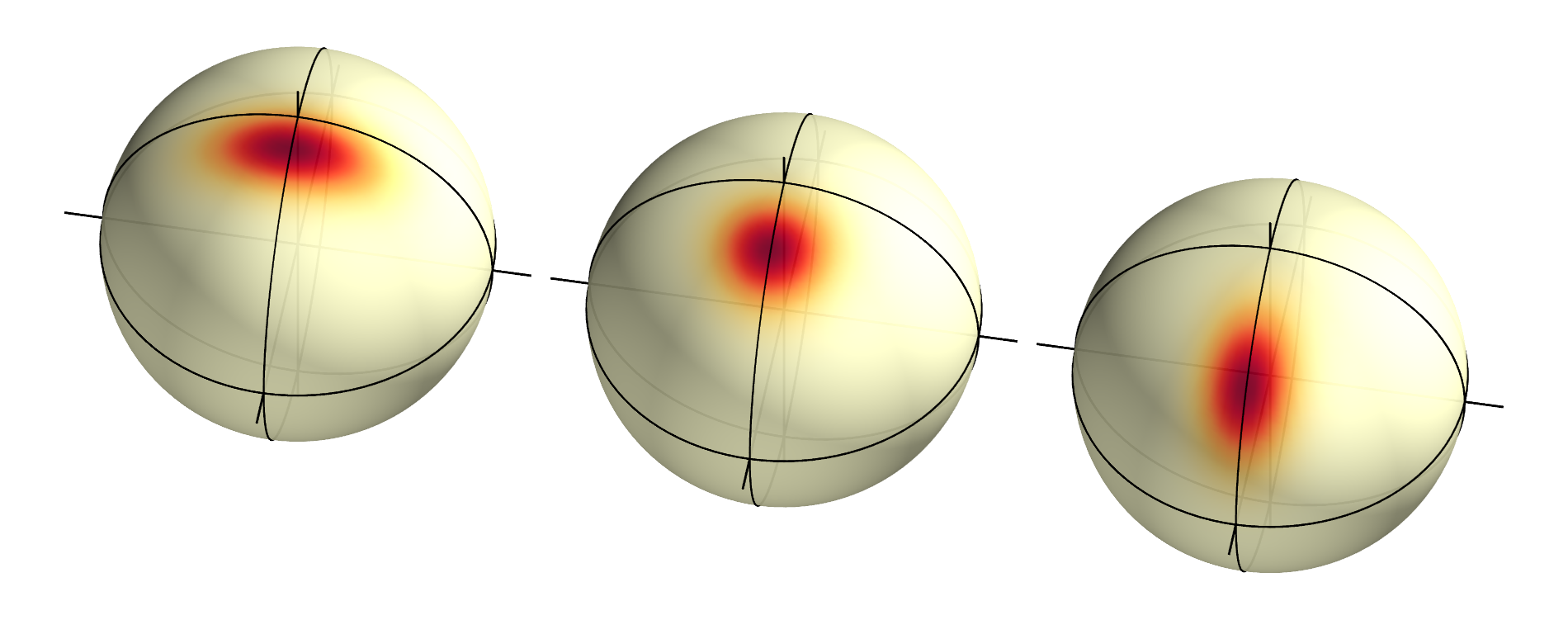}};
	\node[inner sep=0pt] (b) at (0,-1.8)
	    {\includegraphics[width=0.46\textwidth]{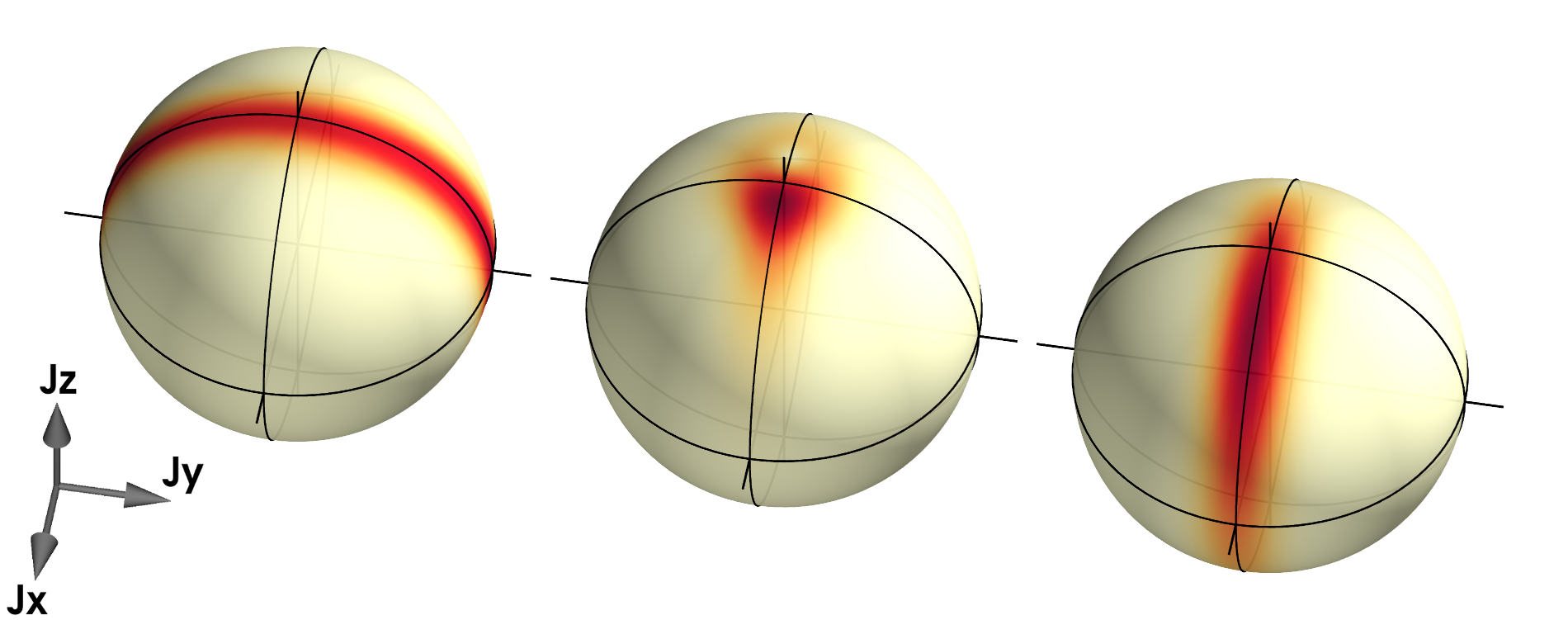}};
	\node[inner sep=0pt] (1) at (-2.6,0.5) {\small \textbf{(i)}};
	\node[inner sep=0pt] (2) at (0,0.) {\small \textbf{(ii)}};
	\node[inner sep=0pt] (2) at (2.6,-0.5) {\small \textbf{(iii)}};
	\end{tikzpicture}
	\caption[Visualization of the TACT echo on the generalized Bloch sphere.]{TACT echo protocol for two dynamical regimes: 
	spin-squeezing (top row) and highly over-squeezed regime (bottom row). Both visualized by the Husimi distribution for $N=100$ atoms.
Top row: (i) Initial spin-squeezing of $-6\,$dB and acquired phase of $\theta = 0.2$. (ii) Ideal inverse squeezing interaction results in a state with 
fidelity of $0.998$ with a CSS. 
(iii) Stronger echo ($r=2$) leads to an anti-squeezed state.
Bottom row: (i) The state possessing the maximum sensitivity reachable with TACT and a phase imprint of $\theta = 0.02$. (ii) Ideal inverse TACT 
(gives a fidelity of $0.814$ with a CSS) and (iii) after an echo with $r=1.5$.
}
		\label{fig2}		
\end{figure}
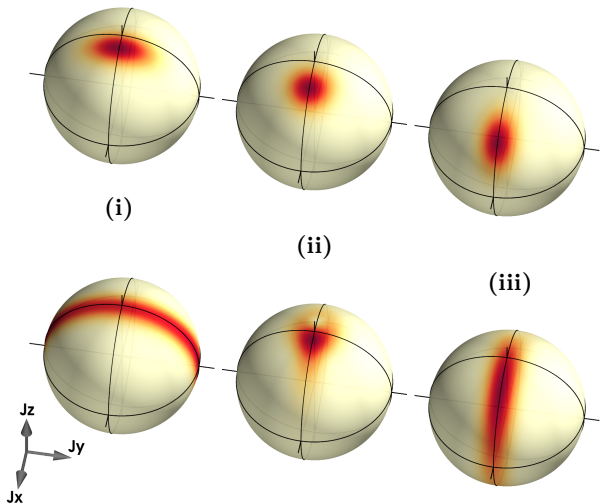

It was shown that the metrological performance produced by TACT dynamics leads to large 
spin-squeezing~\cite{kitagawa1993, Andre2002, Liu2011}, succeeding into a highly entangled state close to a twin-Fock state \cite{Yukawa2014, Kajtoch2015, Zhong2017}.
Figure~\ref{fig2} visualizes the action of the TACT echo protocol for the squeezing regime and also for the optimal---highly entangled---case. 
Both cases will be examined in the following.
Note that the highly entangled state (i, bottom) cannot be detected by measuring first and second moments of the global spin, rendering an exploitation of the entanglement enhancement difficult.
It is a specific advantage of echo schemes that they disentangle the state, returning a fairly classical state to detect.
In this sense, the metrological performance of highly entangled (non-Gaussian) states becomes experimentally accessible \cite{Macri2016}.

\section{One-mode approximation}
The description of the TACT echo can be studied under the assumption that the quantum state remains close to the north pole of the 
Bloch sphere during the protocol (see Fig.~\ref{fig1}).
This assumption is fulfilled for a large number of atoms $N \gg 1$, relatively small squeezing strength $t \chi N \lesssim 1$, and a small interferometric phase shift.
In this case, we can replace the operators $a$ and $a^\dagger$ with the number $\sqrt{N}$ to obtain an effective one-mode description.
The unitary evolution under the TACT Hamiltonian simplifies to the one-mode quadrature-squeezing operator and the interferometric 
rotation becomes the one-mode displacement:
\begin{align}
	e^{-it H_{\rm TACT}} &\rightarrow
	e^{\frac{\gamma}{2} (b^2 - {b^\dagger}^2)} \equiv S(\gamma) \label{squeezing_operator}\\
	e^{-i \theta J_y} &\rightarrow
	e^{\phi ( b^\dagger - b)} \equiv D(\phi)
\end{align}
Here real parameters for the squeezing strength $\gamma = t \chi N$ and the displacement $\phi = \theta \sqrt{N} /2$ arise from the prior two-mode description.

The midfringe position in the two-mode model is on the equator and chosen according to the observable $J_z$. In the one-mode description, the dynamics is confined onto the tangent surface to the Bloch sphere and spanned by the quadratures. 
Therefore, the displacement $\phi$ can be estimated from measurements of the $X = (b + b^\dagger)/2$ quadrature with a phase uncertainty
\begin{align}\label{equ:one_mode_phase_variance}
(\Delta \theta)^2 = \frac{\Delta \left(b + b^\dagger\right)^2}{\left| \partial_{\theta} \braket{b + b^\dagger} \right|^2}.
\end{align}
This can be evaluated analytically for the one-mode equivalent of Eq.~ \eqref{equ:tact_echo_out}, which is $\ket{\psi}_\mathrm{out} = S^{-1}(r \gamma) D(\phi) S(\gamma) \ket{0}$, where the initial coherent state pointing in the $J_z$ direction corresponds 
to all particles in the mode $a$ and thus the vacuum $\ket{0}$ of the mode $b$.
Using textbook formulas (see Appendix for calculations within this section), we find the ideal phase variance to be 
\begin{align}
	\left( \Delta \theta \right)^2 = \frac{e^{-2 \gamma}}{N},
\end{align}
which does not depend on the echo strength and is only determined by the initial state entering the interferometer. 
For positive squeezing strengths $\gamma$, the phase uncertainty $(\Delta \theta)^2$ is decreased below the SQL by a factor of $e^{-2\gamma}$.

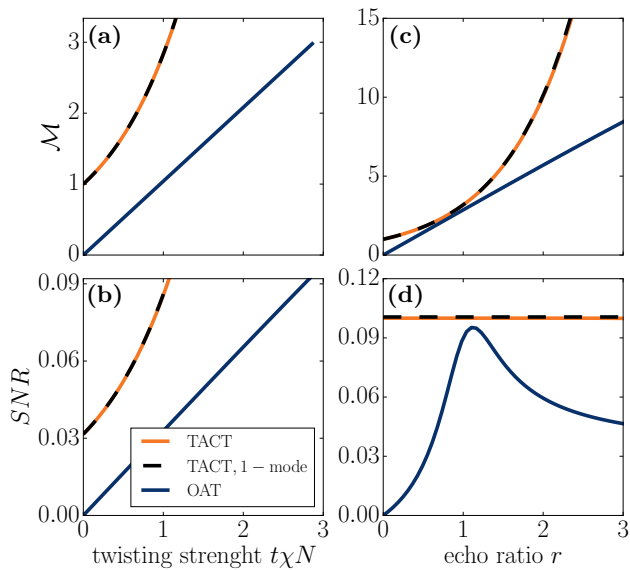
\begin{figure}[t!]
	\centering
  	\begin{tikzpicture}
	\node[inner sep=0pt] (a) at (0,0)
	    {\includegraphics[width=\columnwidth]{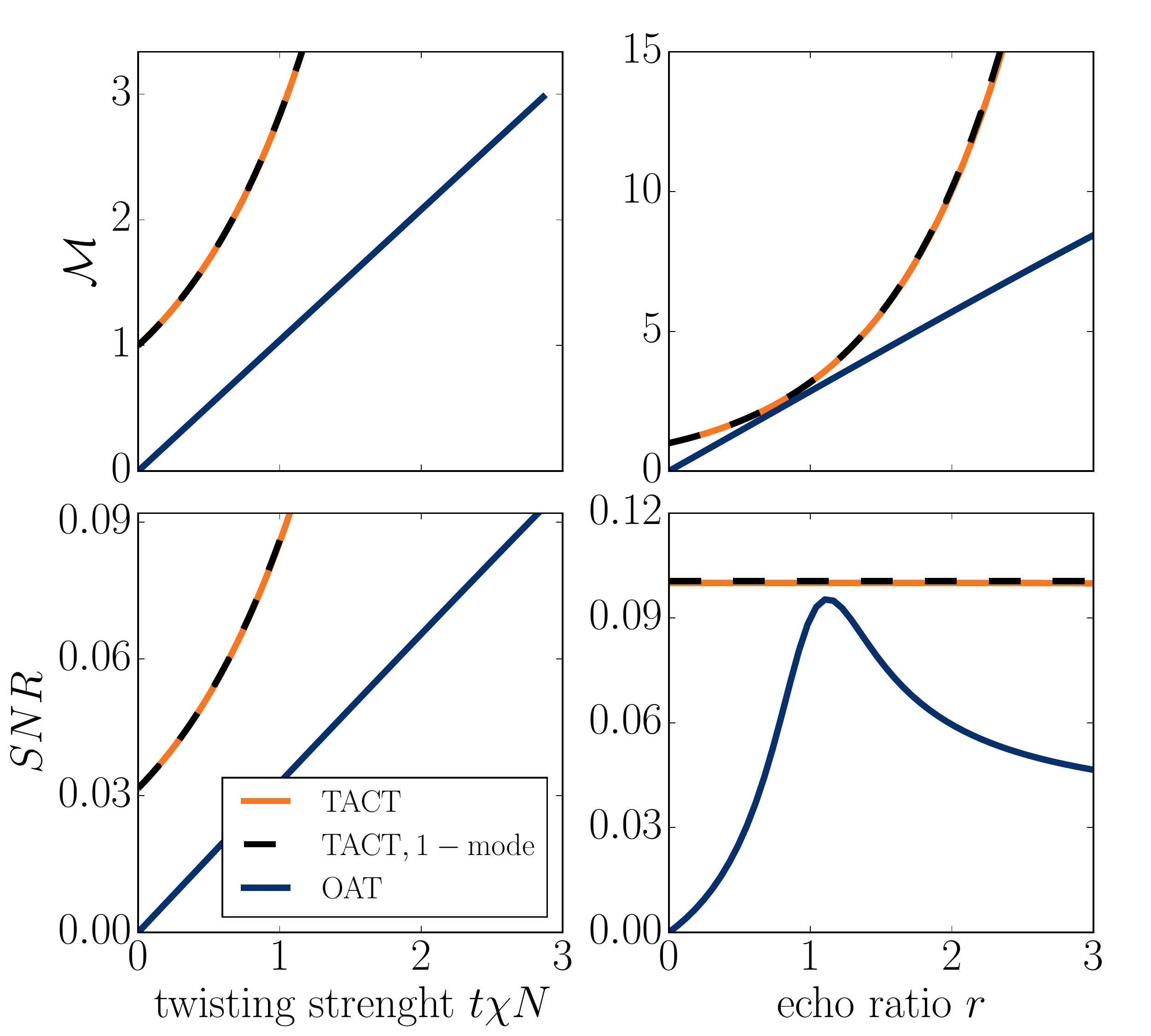}};
	\node[inner sep=0pt] (1) at (-3.0,3.25) {\small \textbf{(a)}};
	\node[inner sep=0pt] (2) at (1.0,3.25) {\small \textbf{(c)}};
	\node[inner sep=0pt] (1) at (-3.0,-0.2) {\small \textbf{(b)}};
	\node[inner sep=0pt] (2) at (1.0,-0.2) {\small \textbf{(d)}};
	\end{tikzpicture}
	\caption[]{(Color online) Comparison of the magnification factor and the signal-to-noise ratio. The orange (light gray) lines show results for the TACT echo and blue (dark gray) lines of the OAT echo respectively. Dashed black lines indicate the analytical one-mode results for the TACT echo. Computations are carried out for $N=10^3$ particles and $\theta=0.001$.
(a, b) Dependence on the interaction strength, showing the range of $0$-$10\,$dB squeezing for both schemes. The echo ratio $r$ is set to $r=1$ in both cases.
(c, d) Dependence on the echo ratio $r$, i.e., the strength of amplification, with initial squeezing of $-10\,$dB. 
}
	\label{fig3}		
\end{figure}

The one-mode model allows for an analytical demonstration of the robustness against detection noise. 
This is implemented by convolving the respective 
distribution of outcomes with a normal distribution of width $\sigma$ \cite{Pezze2013}.
Introducing the influence of detection noise to Eq.~(\ref{equ:one_mode_phase_variance}) is therefore simply accomplished by adding $\sigma^2$ to the nominator, since two convolved Gaussian functions yield a Gaussian function with summed up variances again.
Hence, the noise-dependent phase variance evaluates to
\begin{align}
	\label{equ:noisy_variance}
	( \widetilde{\Delta \theta} )^2 = 
	\left( \Delta \theta \right)^2 + \frac{4 \sigma^2}{N^2 \mathcal{M}^2}.
\end{align}
Here, we defined the magnification factor $\mathcal{M} = \braket{J_x}/\braket{J_x}'$ as the ratio of the signal after and before (primed) the echo.  Within the one-mode approximation it simplifies to 
\begin{align}\label{equ:one_mode_mag_fac}
\mathcal{M} = e^{r \gamma}.
\end{align}
Equation (\ref{equ:noisy_variance}) demonstrates, that the phase magnification strongly suppresses the influence of detection noise.
The simple formula Eq.~(\ref{equ:one_mode_mag_fac}) shows that a stronger echo increases the signal magnification exponentially.
We can of course not conclude that a large magnification cancels the noise influence completely as the depletion of mode $a$ is not captured 
by the one-mode model.
Finally, the signal-to-noise ratio $\mathrm{SNR} \approx \braket{J_x}/\Delta J_x$ is given by
\begin{align}
\mathrm{SNR} = 2\phi e^\gamma,
\end{align} 
revealing the independence of the echo strength.
Thus, the quality of the signal amplification does not decrease for stronger echo interaction.

Figure~\ref{fig3} compares the magnification process of both the OAT and TACT echo scheme in terms of the mentioned quantities.
The TACT echo shows a clear advantage over the OAT scheme.
First, the TACT echo's performance grows exponentially with the interaction time and not just linearly as for the OAT scheme; see Figs.~\ref{fig3}(a)-\ref{fig3}(c).
Furthermore, for our scheme, the signal-to noise ratio is perfectly maintained during the magnification, see Fig.~\ref{fig3}(d), thereby realizing a 
``phase-sensitive noiseless amplification'' previously described in Ref.~\cite{Caves1982}.
In the case of the OAT echo however, the quality of the signal amplification decreases for ratios deviating from the optimal $r \approx 1.1$.
For small twisting strength or echo ratios the measured signal even vanishes completely, since the echo does not yet transfer the signal to the orthogonal readout direction.  
In principle, the results of the OAT scheme can be improved by additional rotations of the quantum state.
We introduce optimizations with respect to rotations before the phase imprint , which optimally align the initial squeezing ellipse, and rotations before the final detection, which correspond to optimizing the readout direction. 
Both optimizations, alone or in combination, can lead to improved results for specific parameter regimes. However, the performance of the OAT echo never surpasses the TACT echo (see Appendix for more details).

Figure~\ref{fig3} also shows that the magnification and the signal-to-noise ratio are well captured by the single-mode approximation.
For over-squeezed states, however, the single mode approximation breaks down and a full two mode analysis is required. 
This is presented in the following sections.

\section{Two-mode model: the squeezing regime}
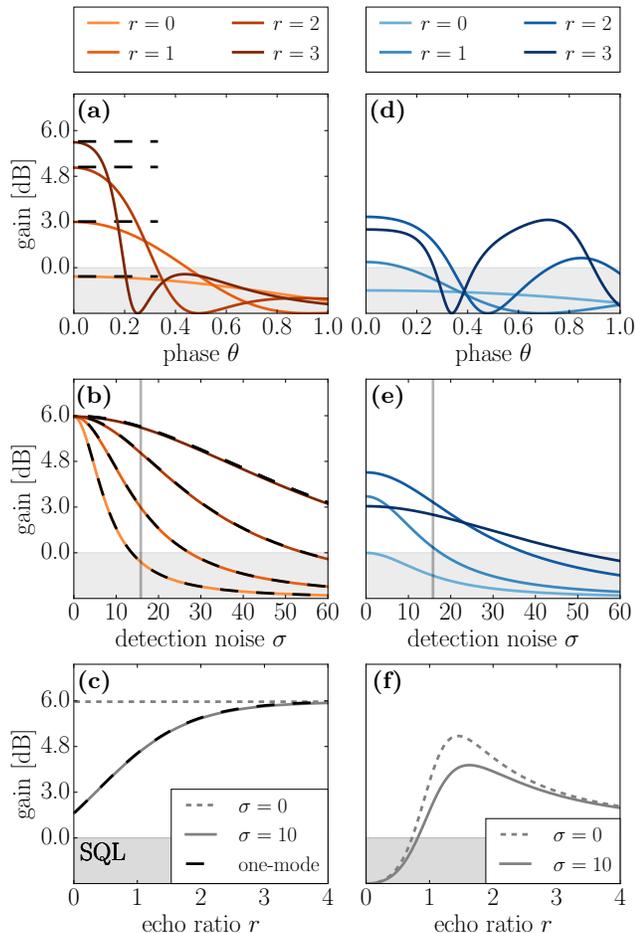
\begin{figure}[t!]
	\centering
	\begin{tikzpicture}
	\node[inner sep=0pt] (left) at (0,0)
	    {\includegraphics[width=\columnwidth]{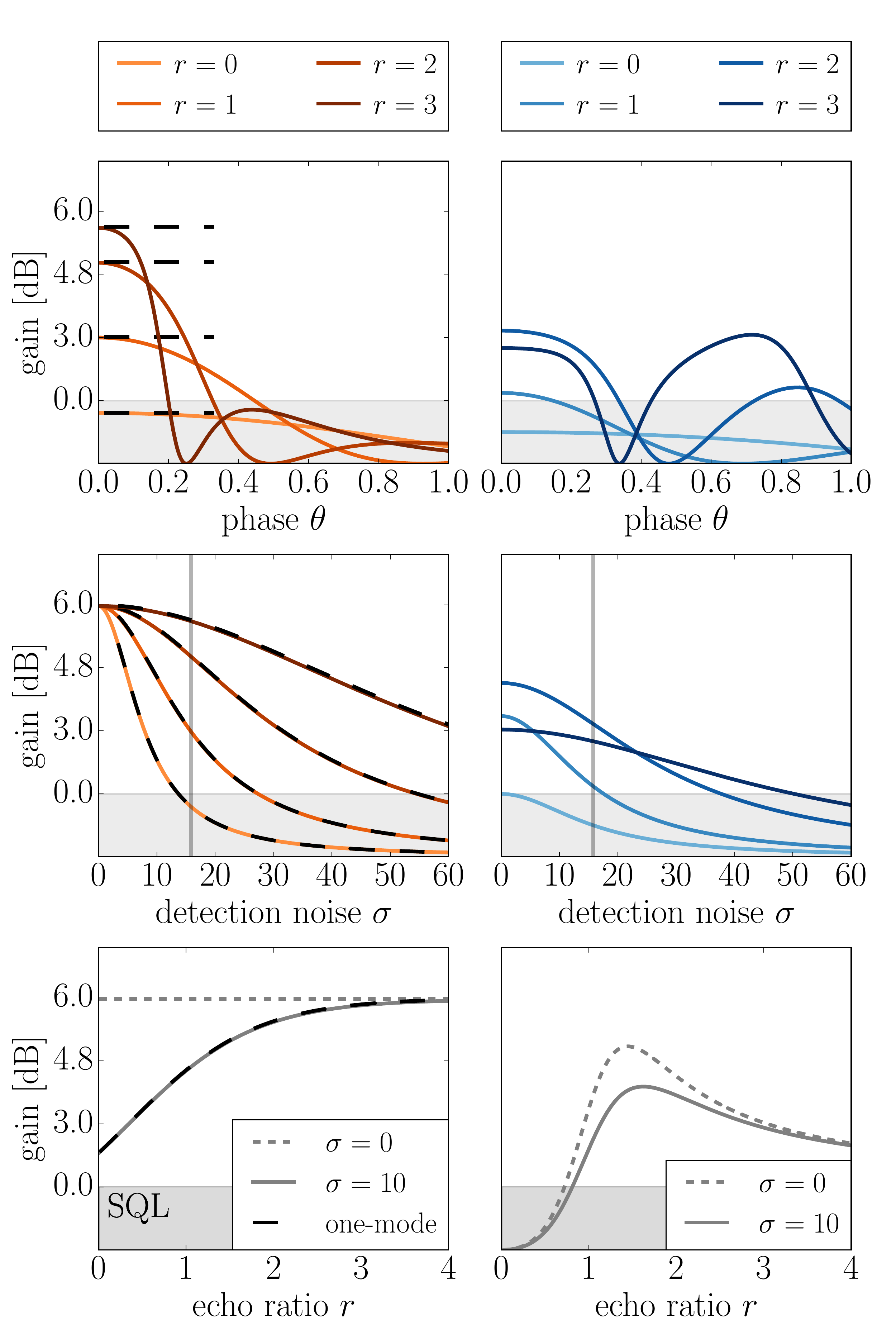}};
	\node[inner sep=0pt] (4) at (-3.1, 4.7) {\small \textbf{(a)}};
	\node[inner sep=0pt] (2) at (-3.1,0.9) {\small \textbf{(b)}};
	\node[inner sep=0pt] (3) at (-3.1,-2.9) {\small \textbf{(c)}};
	\node[inner sep=0pt] (1) at (.8,4.7) {\small \textbf{(d)}};
	\node[inner sep=0pt] (3) at (.8,0.9) {\small \textbf{(e)}};
	\node[inner sep=0pt] (1) at (.8,-2.9) {\small \textbf{(f)}};
	\end{tikzpicture}
	\caption[]{(Color online) Metrological gain of the 
TACT (left panels) and the OAT (right panels) echo schemes, for an initial squeezing of $-6\, \mathrm{dB}$ and a particle number of $N=10^3$. 
Solid colored lines show numerical results for several echo ratios $r$. 
Black dashed lines are the analytical one-mode results $1/( \widetilde{\Delta \theta} )^2 N$ for the TACT echo. 
The shaded area corresponds to a precision within the SQL.
(a,d) Phase dependence of the gain for both echo protocols. The detection noise is set to CSS noise level $\sigma_{\rm CSS} = \sqrt{N}/2$.
(b,e) Noise dependence of the gain. The gray vertical lines indicate the CSS noise level. The phase is set to an optimal small value.
(c,f) Dependence on the echo ratio for the TACT and the OAT echo respectively, for vanishing detection noise (dashed line) and an exemplary noise of $\sigma=10$ (solid line).
}
	\label{fig4}		
\end{figure}
We evaluate the interferometric performance in terms of the classical Fisher information (FI)~\cite{Pezz`e2014, Toth2014, Pezz`e2016}
$F\left(\theta\right) = \sum_\mu \frac{1}{P(\mu \vert \theta)} (\frac{\partial P(\mu \vert \theta)}{\partial \theta})^2$,
where $P(\mu \vert \theta)=\langle \mu \vert \rho_{\rm out}(\theta) \vert \mu \rangle$ is the probability distribution  of the projection over Dicke states $\{\ket{\mu}\}$ with $\mu$ being the eigenvalues of $J_z$.
The FI gives a lower bound on the interferometric phase sensitivity, the Cram\'er-Rao bound $\Delta \theta_{\rm CR} = 1\sqrt{\nu F(\theta)}$, 
for the specific measurement considered, and identifies the metrologically useful entanglement~\cite{Pezze2009}.

For intermediate squeezing strength, the quantum state remains Gaussian and the gain in sensitivity is well captured by the width of its distribution, typically expressed in terms of the spin-squeezing parameter \cite{Wineland1994}
$\xi_R^2 = N \left(\Delta J_x \right)^2 / \braket{J_z}^2$ (here for the case of a mean spin direction along $J_z$ and an interferometer phase shift generated by $J_y$).
In this regime it corresponds to the FI as $F=N/\xi_R^2$.
We choose an initial squeezing of $10 \log_{10}(\xi_R^2)=-6\,$dB as exemplary value, also to allow for experimental resources for an echo ratio larger than one.
Figure~\ref{fig4} shows the metrological gain $10 \log_{10}(F/N)$ as a function of the interferometric phase (top panels) and the detection noise (middle panels).
Here the detection noise is modeled with respect to a measurement of $J_z$.
In Fig.~\ref{fig4} (top), the detection noise is fixed to a level that is equivalent to the quantum noise of a CSS, $\sigma_{\rm CSS} = \sqrt{N}/2$.
Even for such a strong noise contribution, the metrological performance clearly surpasses the SQL and reaches almost the ideal value of $6\,$dB improvement for the case of a strong echo of $r=3$.
In the middle graphs the phase is chosen to be around the optimum $\theta \approx 0$.
The results demonstrate the robust entanglement-enhanced performance under the influence of detection noise. Both echo protocols allow for an estimation of the interferometer signal with sub-SQL precision for noise values well above the CSS
noise level (indicated as gray vertical line).
The proposed TACT echo allows gaining back most of the original sensitivity of the input state and shows better variability regarding the echo ratio, hence it surpasses the results of the OAT echo.
The analytical one-mode results (dashed black lines) agree excellently with the numerical computation in the chosen parameter regime.
Figure~\ref{fig4}~(bottom panels) shows that the OAT scheme demands an optimization of the echo strength for a certain detection noise to gain optimal performance whereas for the TACT echo the larger is the echo ratio $r$, the better; see Fig.~\ref{fig4}(c).
In comparison with Figs.~\ref{fig3}(c) and \ref{fig3}(d), the graphs also illustrate the fact that the SNR reflects the ideal metrological gain, and the magnification factor corresponds to the robustness of the gain to detection noise (the larger $\mathcal{M}$, the more robust).

\section{Optimal performance}
We will now investigate the optimal performance of the TACT echo protocol.
This is achieved for $(t \chi)_{\rm opt} \approx \ln(2 \pi N)/2N$~\cite{Kajtoch2015}, beyond the spin-squeezing regime, 
as obtained from an analysis of the quantum Fisher information (see Appendix).
The corresponding state is shown in Fig.~\ref{fig2}, bottom row (i).
It has a large overlap with a twin-Fock state~\cite{Yukawa2014}, possesses an even higher phase sensitivity as such and thereby outperforms the sensitivity reachable by OAT dynamics on a comparable time scale.

\begin{figure}[t!]
	\centering
  	\begin{tikzpicture}
	\node[inner sep=0pt] (a) at (0,0)
	    {\includegraphics[width=.5\textwidth]{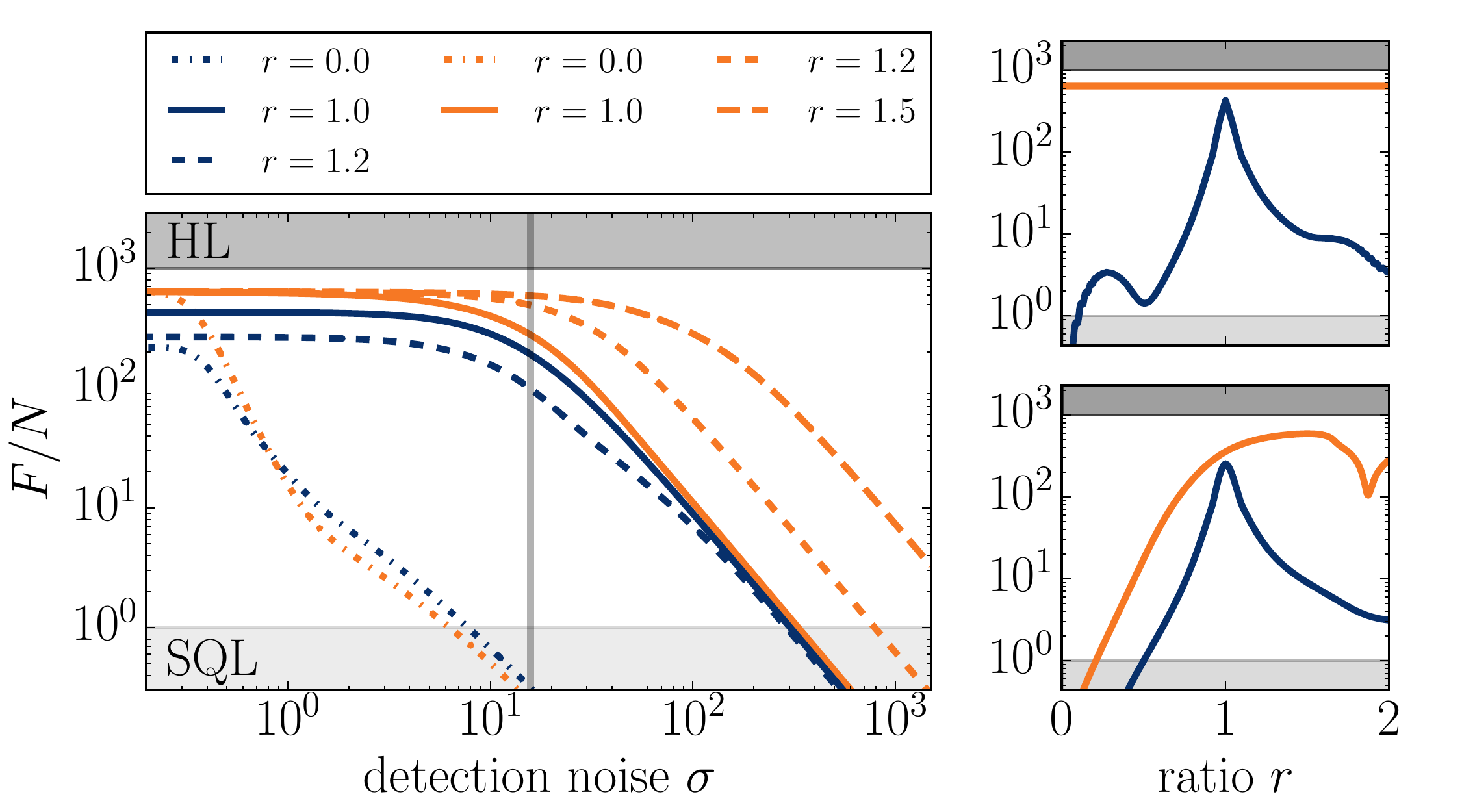}};
	\node[inner sep=0pt] (1) at (.9,0.6) {\small \textbf{(a)}};
	\node[inner sep=0pt] (2) at (2.3,1.7) {\small \textbf{(b)}};
	\node[inner sep=0pt] (2) at (2.3,-.3) {\small \textbf{(c)}};
	\end{tikzpicture}
		\caption{(Color online) Optimal FI achieved with both echo protocols. The orange (light gray) lines show results for the TACT echo. 
		For comparison, we also report the optimal performance of the OAT echo discussed by Davis \textit{et al.}\citep{Davis2016}; see blue (dark gray) lines.
		All computations done for $N=10^3$ atoms.
(a) Dependence on detection noise with an interferometric phase set to a small value which is optimal at CSS noise-level (gray vertical line).
(b) Dependence of the FI on the echo ratio for vanishing detection noise $\sigma=0$.
(c) Same as (b) but for finite detection noise of $\sigma=10$.
Here the phase is set to $\theta=0.002$.
}
		\label{fig5}		
\end{figure}
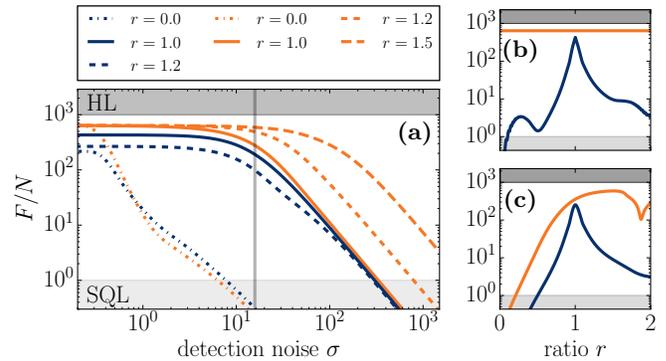
Figure~\ref{fig5}(a) shows the achievable FI in this regime under the influence of detection noise. 
Without echo ($r=0$, dashed-dotted lines), the performance already starts to decline at $\sigma \approx 0.3$ 
since an operation close to the Heisenberg limit typically requires single-atom resolving detection.
This requirement is avoided by both the OAT~\cite{Davis2016} and the TACT echo protocol, maintaining Heisenberg-limited sensitivity for detection noise levels far beyond what has been demonstrated experimentally in several groups.
A specific advantage of the TACT echo scheme, the nonessential reversal of the twisting dynamics ($r=1$), holds even in this highly entangled regime.
For vanishing detection noise, the TACT echo is independent of the echo ratio whereas the OAT echo has a sharp optimal performance at $r=1$; see Fig.~\ref{fig5}(b).
For nonvanishing detection noise (even for detection noise around the CSS noise level) 
the TACT echo still shows a fairly constant performance in a broad range of echo ratios, see Fig.~\ref{fig5}(c).

\section{Asymptotic scaling for a large number of atoms}
\begin{figure}[t!]
	\centering
	\includegraphics[width=\columnwidth]{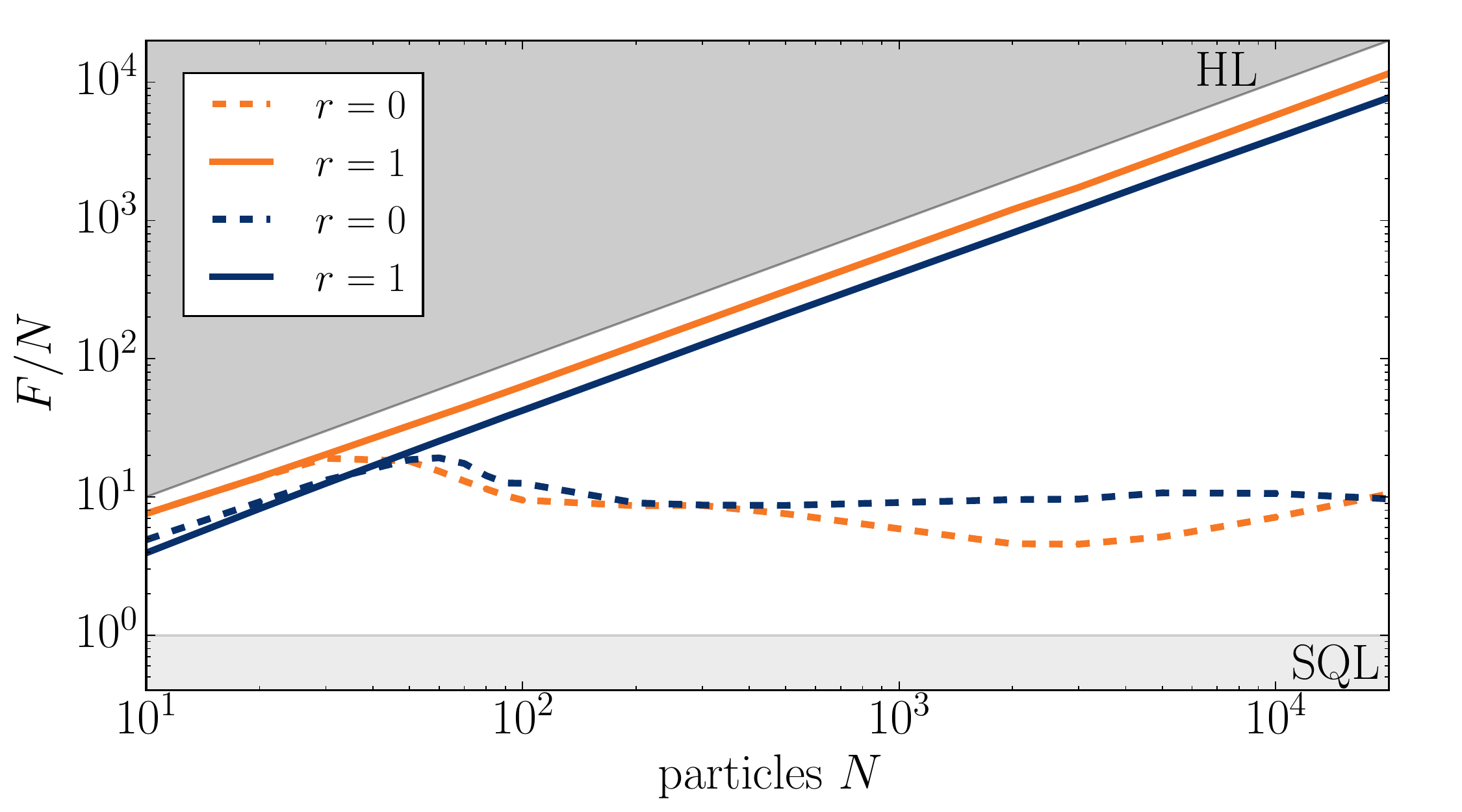}
		\caption[]{(Color online) FI as a function of the atom number for both echo protocols (solid lines) in the presence of a detection noise $\sigma = \sqrt{N}/10$. The orange (light gray) and blue (dark gray) lines show the results of the TACT and the OAT echo respectively. The echo clearly retains the Heisenberg scaling. For comparison, the dashed lines show the resulting FI without echo and in presence of the same number-dependent detection noise. In this case, the Heisenberg limit is quickly lost and the sensitivity follows a scaling $\sqrt{N}$, with an improvement over the SQL (gray area) 
		given only by a constant factor for large $N$.}
		\label{fig6}		
\end{figure}

It has been emphasized in the literature that, in the presence of a variety of single-particle (uncorrelated) decoherence sources,
the quantum-enhancement cannot maintain the Heisenberg scaling asymptotically in the number of particles.
In the limit $N \rightarrow \infty$, the sensitivity scales with the square root of the particle number, eventually reaching only a constant factor beyond the SQL~\cite{Escher2011, Demkowicz2012}.
Accordingly, our numerical results show that in the case without echo the Heisenberg scaling is lost quickly and SQL scaling is obtained for large $N$.
However, exploiting an interaction-based readout within
our TACT echo scheme---as well as with the OAT echo---the Heisenberg scaling is preserved in the presence of a large detection noise and atom number (up to $N \sim 10^4$).
Figure~\ref{fig6} shows the FI with an exemplary detection noise of $\sigma(N) = \sqrt{N}/10$, 
which scales with the square root of the number of atoms (such as the photon shot noise in fluorescence detection).
The apparent contradiction with the mentioned no-go results is explained by the fact that 
interaction-based detection cannot be modeled as a single-particle noisy channel and does not fall into 
the range of noisy interferometers analyzed in Refs.~\cite{Escher2011, Demkowicz2012}.
Besides our findings, asymptotic scaling that surpasses the SQL was also reported for non-Markovian phase noise \cite{Chin2012} 
and noise that is perpendicular to the phase imprint \cite{Chaves2013, Brask2015}.
Note that in the case of detection noise considered here, the noise is parallel to the parameter to be estimated.

While our results are obtained numerically and cannot be evaluated in the limit of infinite atom numbers, 
we argue that they correctly reflect the asymptotic behavior.
Nevertheless, we would like to note that the asymptotic scaling of the sensitivity with the number of particles is typically not relevant in actual metrological experiments.  
Besides the noise that scales with the square root of the particle number,
there are further technical noise components that are linear in the particle number and take over while increasing $N$ (see for example Ref.~\cite{Hume2013}).
These noise contributions eventually destroy both the Heisenberg and the SQL scaling, yielding an interferometric precision that is independent of the particle number.

\section{Implementation in spinor Bose-Einstein condensates}

Bose-Einstein condensates (BECs) with a spin degree of freedom---in our case spin $F=1$---undergo spin-changing collisions, 
creating entangled atom pairs and many-particle entanglement of indistinguishable atoms~\cite{Duan2000, DuanPRA2002,Pu2000, Pezz`e2016}. 
This type of entanglement generation has been demonstrated by several groups~\cite{Gross2011, Luecke2011, Hamley2012, Pezz`e2016, Luo2017}.
In particular, states generated by spin-changing collisions have been applied for the demonstration of 
Einstein-Podolsky-Rosen entanglement~\cite{Peise2015a}, interaction-free measurements~\cite{Peise2015}, 
interferometric applications~\cite{Luecke2011,Kruse2016,Linnemann2016}, and---most recently---for the generation of spatially 
distributed entanglement~\cite{Kunkel2018, Fadel2018, Lange2018}.
We will now show that spin dynamics in atomic BECs allows for an effective realization of the TACT echo protocol in the spin-squeezing regime.

In the following we consider a BEC initially prepared in $m_F=0$.
The evolution under spin changing collisions is described by a four-wave mixing Hamiltonian 
$H_{\rm FWM} = 2 \lambda (  a^\dagger_{0} a^\dagger_{0} a_{1} a_{-1} + a^\dagger_{1} a^\dagger_{-1} a_{0} a_{0})$~\cite{Law1998, Stamper-Kurn2013}. 
Additional spin-preserving collisions are described by the collisional shift $H_{\rm CS} = \lambda(2\hat{N}_0 -1)(\hat{N}_1 + \hat{N}_{-1})$.
Here, the coupling strength $\lambda$ is determined by the scattering lengths of allowed channels and by the spatial modes of the atoms.
Furthermore, the term $H_{\rm QZ} = q(\hat{N}_1 + \hat{N}_{-1})$ includes the quadratic Zeeman shift $q \propto B^2$ that depends on the external magnetic field strength $B$.
Alternatively, $q$ can be experimentally tuned by coupling the Zeeman levels to a different hyperfine level by an adjustable microwave dressing field.
By introducing a change of basis to the symmetric (S) and the antisymmetric (A) combinations of the $m_F = \pm 1$ states,
\begin{align}
a_S = \frac{a_{+1} + a_{-1}}{\sqrt{2}} \quad \text{and} \quad
a_A = \frac{a_{+1} - a_{-1}}{\sqrt{2}},
\end{align}
the overall Hamiltonian $H = H_{\rm QZ} + H_{\rm CS} + H_{\rm FWM}$ becomes
\begin{align}
H&=q (\hat{N}_S + \hat{N}_A ) \label{equ:ham_a}\\
&+ \lambda (2\hat{N}_0 -1 ) (\hat{N}_S + \hat{N}_A )\label{equ:ham_b}\\
&+ 2 \lambda [ ( J_{x,S}^2 - J_{y,S}^2 ) - ( J_{x,A}^2 - J_{y,A}^2 ) ],\label{equ:ham_c}
\end{align}
where $J_{x,S} = (a_0^\dag a_{S} + a_{S}^\dag a_0)/2$, $J_{y,S} = (a_0^\dag a_{S} - a_{S}^\dag a_0)/2i$, 
$J_{x,A} = (a_0^\dag a_{A} + a_{A}^\dag a_0)/2$ and $J_{y,A} = (a_0^\dag a_{A} - a_{A}^\dag a_0)/2i$.
In the limit of a large number of atoms in $m_f=0$, the evolution in the three modes can thus be described within two SU(2) systems $(0,A)$ and $(0,S)$, with the corresponding pseudo-spin operators $J_{i,A}$ and $J_{i,S}$ for $i=x,y,z$.
The four-wave mixing part of the Hamiltonian realizes the TACT in the symmetric and the antisymmetric mode respectively. 
This description is valid for comparably short interaction times, where 
$N_0$ can be assumed constant and
the Zeeman term Eq.~(\ref{equ:ham_a}) can be tuned to cancel the collisional shift Eq.~(\ref{equ:ham_b}).
For long interaction times, depletion of the initial condensate and quantum fluctuations become relevant:
the terms Eqs.~(\ref{equ:ham_a}) and (\ref{equ:ham_b}) do not cancel perfectly and the evolution deviates from pure TACT dynamics.
Within the two-mode system $(0,S)$, it is possible to realize a two-mode Ramsey-like interferometric protocol~\cite{Kruse2016}:
Beam splitters can be realized by resonant radio-frequency coupling, which does not couple to the antisymmetric mode $(A)$.
Interferometric phase shifts according to $\exp (-i\theta J_{y,S})$ can be engineered for typical applications, 
such as the measurement of magnetic fields or, as demonstrated in Ref.~\cite{Kruse2016}, the measurement of time.

Viewed in the original basis $m_F = 0, \pm 1$, this implementation of the TACT echo is similar to the pumped-up-SU(1,1) interferometer \cite{Szigeti2017}
where a linear coupling of the atoms in $m_F = 0$ to the side modes (realizing a ``tritter") 
happens before and after the interferometric phase imprint on the side modes.
The linear coupling corresponds to the beam splitter between $m_F = 0$ and the symmetric mode and the phase on the side modes is equivalent to a rotation by $J_z^{(0,S)}$.  Therefore, our phase imprint $\exp (-i\theta J_{y,S})$, combined with the spin dynamics in the preparation and in the detection stage, resembles the mentioned scheme.

Within the two-mode system (0,S), the four-wave mixing Hamiltonian simplifies to the TACT Hamiltonian Eq.~(\ref{equ:H_TACT}).
The only difference is that the squeezing and anti-squeezing directions are rotated in the $x$-$y$ plane by $\pi/4$, which can be easily compensated by an additional phase shift after the interaction [this corresponds to optimizing the tritter parameters in the pumped-up SU(1,1) scheme].
The interferometer only couples to the antisymmetric mode $(A)$ by the four-wave mixing Hamiltonian. However, for moderate squeezing strengths, the number of atoms in the mode $(A)$ remains small.
It influences the interferometer signal only as a small loss of contrast, as the atoms do not sense the relative phase shift.
Regarding the output observable, it is difficult to detect the number of atoms in the modes $(S)$ and $(A)$ independently.
Thus, we consider the inferior but more practical measurement of the total number of atoms in both modes $(S)$ and $(A)$.
This total number $\hat{N}_{S} + \hat{N}_{A} = \hat{N}_{+1} + \hat{N}_{-1}$ approximates a direct measurement of $\hat{N}_{S}\approx \hat{N}_{S} + \hat{N}_{A}$, 
as an initial population of mode (A) by the TACT interaction is mostly removed during the echo (regarding $r \approx 1$).
Since the sign of the coupling strength $\lambda$ cannot be inverted, the state must be rotated by $\pi/2$ to achieve inverse squeezing during the echo.
In the limit of strong squeezing, the terms Eqs.~(\ref{equ:ham_a}) and (\ref{equ:ham_b}) do not cancel perfectly, 
and the echo does not constitute a perfect time reversal.
However, in the case of initial compensation, i.e. $q=-\lambda(2N-1)$, the dynamics resembles a clean TACT echo up to relatively large squeezing values.

\begin{figure}[t!]
	\centering
  	\begin{tikzpicture}
	\node[inner sep=0pt] (a) at (0,0)
	    {\includegraphics[width=.5\textwidth]{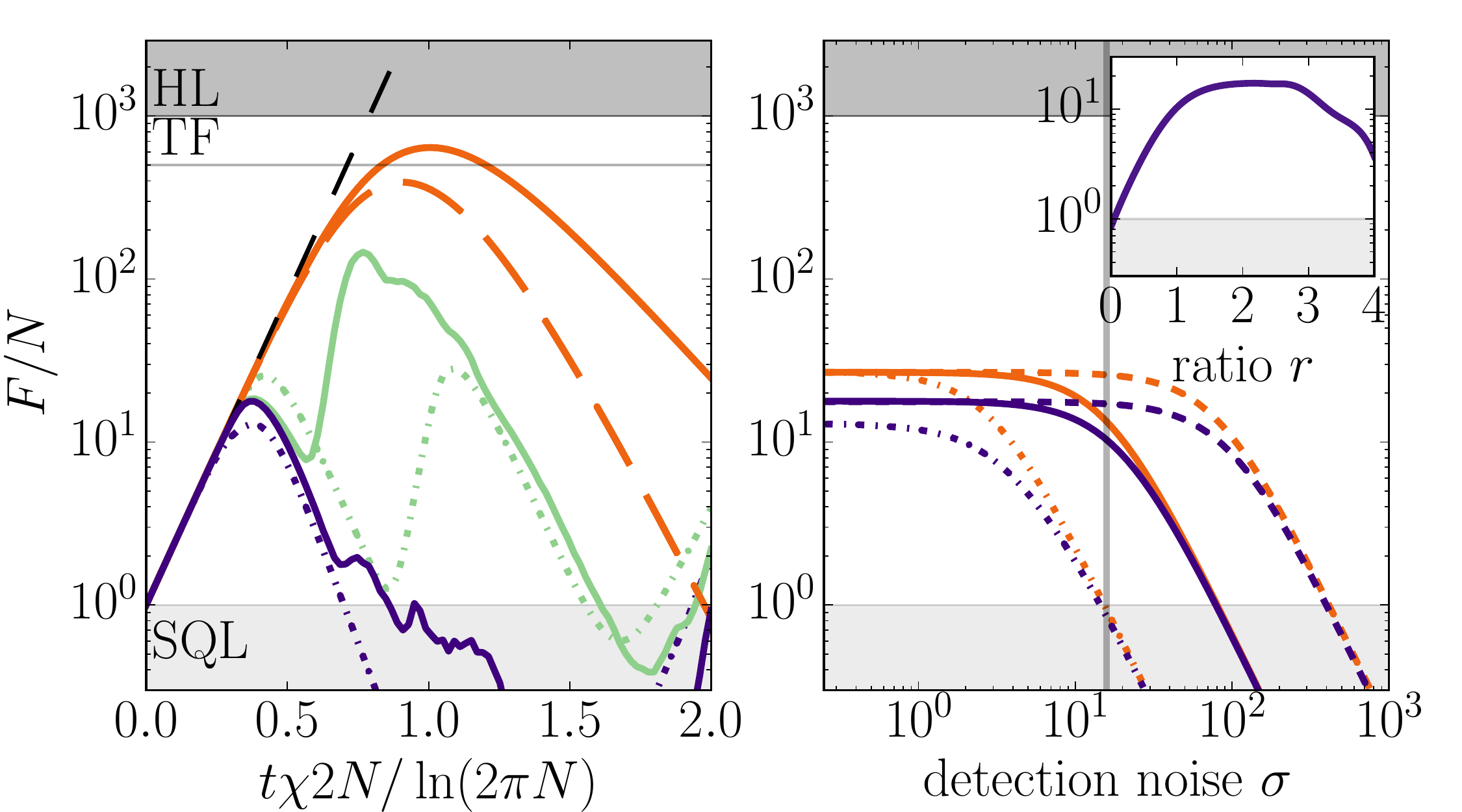}};
	\node[inner sep=0pt] (1) at (-3.9,2.2) {\small \textbf{(a)}};
	\node[inner sep=0pt] (2) at (.25,2.2) {\small \textbf{(b)}};
	\end{tikzpicture}
	\caption[]{(Color online) The FI achieved by the TACT echo as implemented with a spinor BEC. The purple (dark gray) lines show a realistic experimental scenario, which is affected with respect to the ideal two-mode TACT scheme, see orange (medium gray) lines, by three causes. First, TACT occurring in the symmetric and the antisymmetric mode [that is evolution by (\ref{equ:ham_c}) only] depicted by the orange (medium gray) dashed line. Second, the additional imperfect canceling of the collisional shift [that is evolution by $H$ and measurement of the number of particles in $(S)$ and $(A)$ separately] depicted by the green (light gray) lines. Third, the additional realistic measurement depicted by the purple (dark gray) lines.
The dashed black line shows the analytical one-mode results. Computations carried out for $N=10^3$ atoms and an interferometric phase of $\theta = 0.002$.
(a) Twisting strength dependence for $r=0$ (dashed-dotted) and $r=1$ (solid) and vanishing detection noise.
(b) Robustness against detection noise for $r=0$ (dashed-dotted), $r=1$ (solid), and $r=2$ (dashed). Both, the clean TACT echo and the realistic implementation with spin dynamics, evaluated at the point where the purple (dark gray) line in (b) reaches its maximum. The inset shows the dependence on the echo ratio for the realistic scenario with a detection at the CSS noise level.}
		\label{fig7}		
\end{figure}

To investigate the applicability of the TACT echo in spinor BECs, we have performed 
numerical simulations of the full three-mode evolution.
Figure \ref{fig7} shows the realistic performance of the echo implemented with spin-changing collisions in a spinor BEC (purple lines).
The results include the effects of loss to the antisymmetric mode, the effect of the 
terms Eqs.~(\ref{equ:ham_a}) and (\ref{equ:ham_b}), and the measurement of $N_S+N_A$ discussed above.
Figure \ref{fig7} also quantifies the relative strength of these unwanted contributions, 
which have been obtained by considering only the corresponding parts of the spin dynamics Hamiltonian.
The green lines show the FI obtained when $N_S$ and $N_A$ are measured separately (see Appendix for details).
The gray line only takes into account the effect of the antisymmetric mode being populated, that is time evolution by Eq.~(\ref{equ:ham_c}) only.
Hence, the implementation by spin dynamics is particularly hampered by the effect of the additional terms Eqs.~(\ref{equ:ham_a}) and (\ref{equ:ham_b}) in the Hamiltonian.
Here, the ideal FI is not fully independent of the echo, therefore we show results for $r=0$ (dashed-dotted lines) and $r=1$ (solid lines).
In the presented case of $N=10^3$ atoms, spin dynamics perfectly reproduces the TACT echo for up to $12.5\,$dB of quantum-enhancement.
Note that the optimal gain over the SQL will not remain constant for larger particle numbers, but is expected to scale with the number of employed atoms 
(see  Fig.~\ref{fig8} in Appendix), allowing for even stronger squeezing at typical atom numbers in the $10^4$ range.
Figure~\ref{fig7}~(b) further shows that the echo implemented via spin dynamics features the same robustness against detection noise as the clean TACT scheme.
The interaction-based readout based on this echo protocol allows to retain the initial quantum-enhancement with a detection at CSS noise level.

\section{Conclusion}
In summary, we have demonstrated an interaction-based readout interferometric scheme that exploits TACT dynamics and 
largely overcomes the limitations imposed by detection noise. 
The TACT echo protocol provides superior performances with respect to the OAT echo scheme in both, the spin squeezing and the highly over-squeezed regime.
In particular, both echo protocols allow for 
a Heisenberg scaling of the interferometric sensitivity in the limit of large particle numbers and with number-dependent detection noise present.
There exist several possibilities for implementing the TACT Hamiltonian in realistic physical systems, 
including a recent proposal for atom-light interaction in a cavity~\cite{Zhang2017}.
We have studied here the realization of the TACT echo protocol using spin dynamics in spinor BECs and highlighted the analogy to pumped-up SU(1,1) interferometry. 
In the future, we aim at the experimental demonstration of the TACT echo in a microwave clock.

\begin{acknowledgments}
We thank Monika Schleier-Smith for helpful discussions and Marco Gabbrielli for stimulating remarks and for reading the manuscript. We also thank an anonymous referee for important remarks that triggered further analysis.
C.K. acknowledges support from the Deutsche Forschungsgemeinschaft through CRC 1227 (DQ-mat), Project No. A02.
F.A.
acknowledges support from the Hannover School for Nanotechnology (HSN) and the ERASMUS+ program.
\end{acknowledgments}

\appendix
\section{One-mode calculations}
We derive the one-mode formulation of the two-mode phase variance from the method of moments formula. Including the detection-noise model as explained in the main text, the noisy phase variance reads
\begin{align}
(\widetilde{\Delta \theta})^2 = \frac{(\Delta J_x)^2 + \sigma^2}{|\partial_\theta \braket{J_x}|^2},
\end{align}
with $(\Delta J_x)^2 = \braket{J_x^2} - \braket{J_x}^2$.
Replacing
\begin{align}
\braket{J_x} = \frac{1}{2} \braket{a^\dagger b + a b^\dagger}
&\rightarrow \frac{\sqrt{N}}{2} \braket{b^\dagger + b},
\end{align}
the one-mode approximation of the above formula reads
\begin{align}
(\widetilde{\Delta \theta})^2 &= \frac{\braket{(b^\dagger)^2 + b^2 + 2b^\dagger b +1}-\braket{b^\dagger + b}^2}{|\partial_\theta \braket{b^\dagger +b}|^2} \nonumber \\ 
&+ \frac{4 \sigma^2}{N|\partial_\theta \braket{b^\dagger +b}|^2}.\label{equ:noisyPV}
\end{align}
Relevant states within the one-mode approximation are the state before the echo $\ket{\psi}' = D(\phi) S(\gamma) \ket{0}$ and the output state after the echo $\ket{\psi}_\mathrm{out} = S^{-1}(r \gamma) D(\phi) S(\gamma) \ket{0}$, with the displacement and the squeezing operator defined as in the main text.
Applying textbook properties of these operators (see, for example, Ref.~\cite{ScullyBOOK}), we evaluate the following expectation values:
\begin{align*}
\braket{b^\dagger}' &= \bra{0} S^\dagger D^\dagger b^\dagger D S \ket{0} \\
&= \bra{0} S^\dagger (b^\dagger + \phi) S \ket{0} = \phi, \\
\braket{b^\dagger} &= \bra{0} S(\gamma)^\dagger D^\dagger {S(r\gamma)^{-1}}^\dagger b^\dagger S(r\gamma)^{-1} D S(\gamma) \ket{0} \\
&= \bra{0} S(\gamma)^\dagger \left( (b^\dagger + \phi) \cosh(r \gamma) \right.
\\
&\quad + \left. (b + \phi) \sinh(r \gamma) \right) S(\gamma) \ket{0}\\
&= \phi e^{r \gamma}
\end{align*}
The expectation values $\braket{b}' = \phi$ and $\braket{b} = \phi e^{r\gamma}$ are calculated analogously.
With this the magnification factor becomes
\begin{align}
M_{\rm TACT} &= \frac{\braket{J_x}}{\braket{J_x}'} = e^{r\gamma}.
\end{align}

Introducing the following shorthand notation
\begin{align}
S_1 &= S(\gamma), \quad
\mu_1 = \cosh(\gamma), \quad
\nu_1 = \sinh(\gamma) \\
S_2 &= S(r \gamma), \quad
\mu_2 = \cosh(r \gamma), \quad
\nu_2 = \sinh(r \gamma)
\end{align}
we evaluate the relevant expectation values for $\ket{\psi}_\mathrm{out}$  as follows.
\begin{widetext}
\begin{align*}
\braket{b^\dagger b}
&= \bra{0} S_1^\dagger D^{-1} S_2 b^\dagger  b S_2^\dagger D S_1 \ket{0} \\
&= \bra{0} ( (b^\dagger \mu_1 - b \nu_1 + \phi) \mu_2 + (b \mu_1 - b^\dagger \nu_1 + \phi) \nu_2) ( (b \mu_1 - b^\dagger \nu_1 + \phi) \mu_2 + ( b^\dagger\mu_1 - b\nu_1 + \phi ) \nu_2) \ket{0}\\
&= (\mu_1 \nu_2 - \nu_1 \mu_2)^2 + \phi^2 (\mu_2 + \nu_2)^2
\\
\\
\braket{(b^\dagger)^2}
&= \braket{0|S^\dagger_1 D^\dagger S_2 b^\dagger S^{-1}_2 S_2 b^\dagger S^{-1}_2 D S_1 |0}  \\
&= \braket{0| [(b^\dagger \mu_1 - b\nu_1 + \phi)\mu_2 + (b\mu_1 - b^\dagger\nu_1 + \phi )\nu_2 ]^2|0} \\
&= (\phi^2 - \mu_1 \nu_1) (\mu_2^2 + \nu_2^2) + (\mu_1^2 + \nu_1^2 + 2\phi^2)\mu_2 \nu_2
\end{align*}
The remaining expectation value is equal to the above calculated $\braket{b^2} = \braket{(b^\dagger)^2}$.
For the denominator of the phase variance we find $|\partial_\theta \braket{b^\dagger +b}|^2 = N \exp(2r\gamma)$, determining the noise dependent term of Eq.~(\ref{equ:noisyPV}). The ideal phase variance evaluates to:
\begin{align*}
(\Delta \theta)^2
&= \frac{2 (\phi^2 - \mu_1\nu_1) (\mu_2^2 + \mu_2^2) + 2(\mu_1^2 + \nu_1^2 + 2 \phi^2)\mu_2 \nu_2 + 2(\mu_1\nu_2 - \nu_1 \mu_2)^2 + 2\phi^2 (\mu_2 + \nu_2)^2 +1 - 4 \phi^2 e^{2 r \gamma}}{N e^{2 r \gamma}} \\
&= \frac{e^{2\gamma(r-1)}(1+4 \phi^2e^{2\gamma}) - 4 \phi^2 e^{2r\gamma}}{N e^{2 r \gamma}}\\
&= \frac{e^{2\gamma(r-1)}}{N e^{2 r \gamma}}\\
&= \frac{e^{-2\gamma}}{N}.
\end{align*}
\end{widetext}

\section{The optimal parameters}
\begin{figure}[h!]
	\centering
	\includegraphics[width=.45\textwidth]{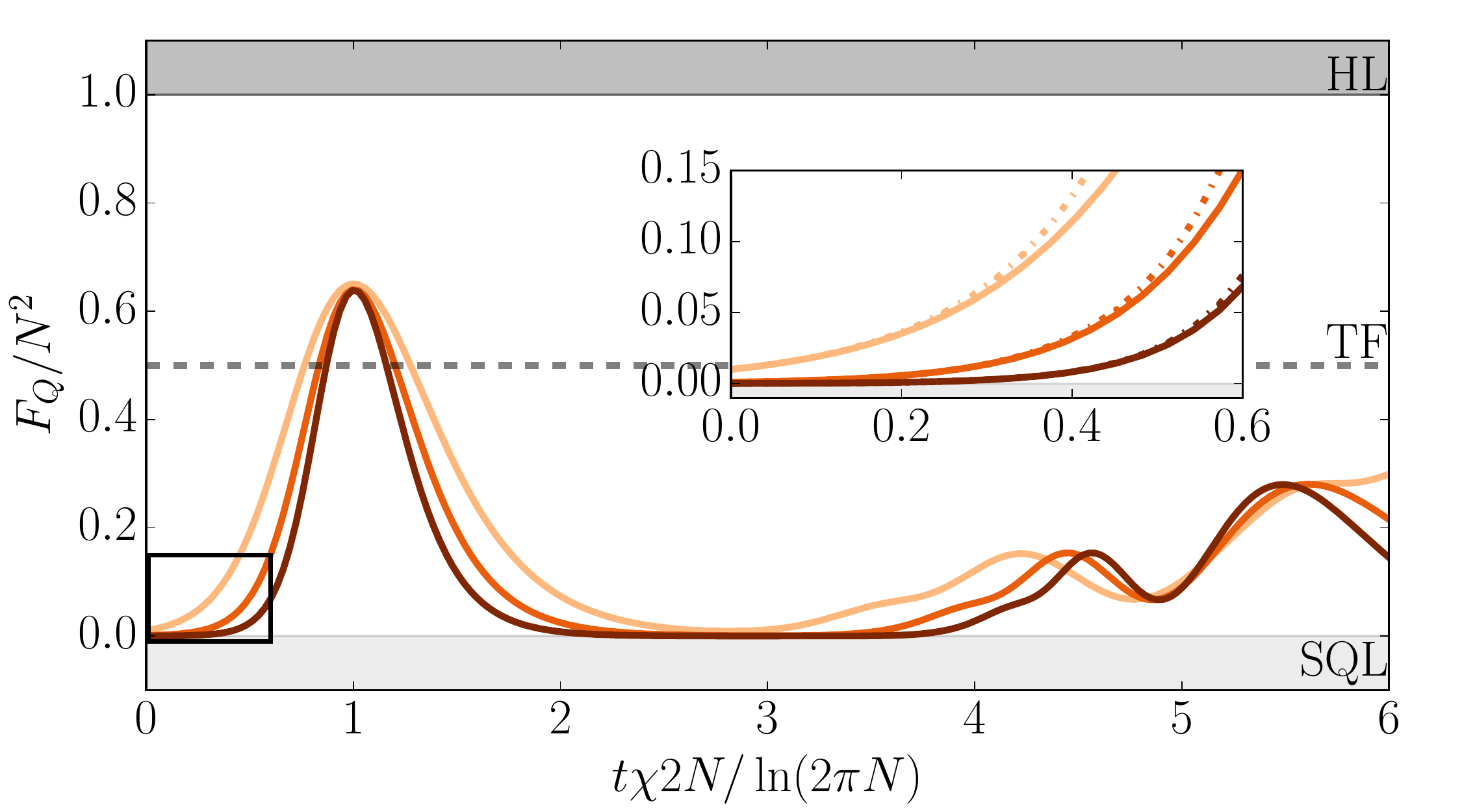}
		\caption{(Color online) The QFI achieved by TACT for $N=10^2, 10^3, 10^4$ (from bright to dark color) depending on the twisting strength. Scaling of the axes are chosen to overlie the maxima for different number of particles. The inset shows the range where the one-mode results coincide with the numerical two-mode computations. The models deviate  about $5\%$ at a spin-squeezing parameter of $\xi_\mathrm{R}^2 = -7.6, -15.8, -25.8\, \mathrm{dB}$, respectively.}
		\label{fig8}		
\end{figure}

The parameters for the generation of an optimally entangled state by TACT can be obtained for arbitrary particle numbers employing the scaling behavior reported in Ref.~\cite{Kajtoch2015}.
Figure~\ref{fig8} shows the quantum Fisher information (QFI) as a function of the scaled evolution time.
The maximum QFI corresponds to the optimal value for the TACT echo and exceeds the performance reached by the twin-Fock state (TF) which is also the value of the characteristic plateau the OAT dynamics reaches \cite{Davis2016}.

\section{Optimization of the OAT echo}
\begin{figure}[t!]
	\centering
  	\begin{tikzpicture}
	\node[inner sep=0pt] (a) at (0,0)
	    {\includegraphics[width=\columnwidth]{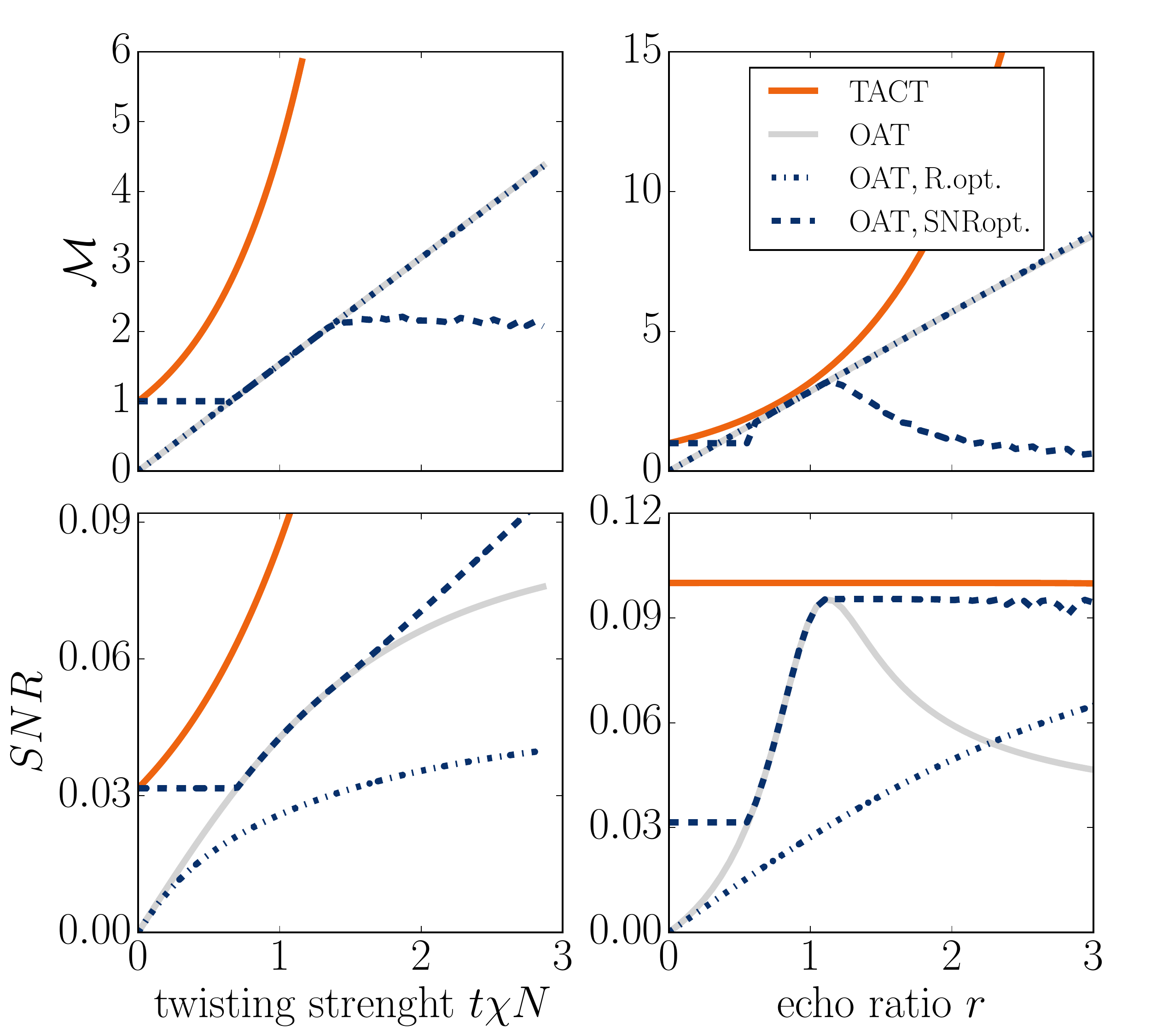}};
	\node[inner sep=0pt] (1) at (-3.0,3.25) {\small \textbf{(a)}};
	\node[inner sep=0pt] (2) at (1.0,3.25) {\small \textbf{(c)}};
	\node[inner sep=0pt] (1) at (-3.0,-0.2) {\small \textbf{(b)}};
	\node[inner sep=0pt] (2) at (1.0,-0.2) {\small \textbf{(d)}};
	\end{tikzpicture}
	\caption[]{(Color online) The magnification factor and the signal-to-noise ratio for two relevant optimizations of the OAT echo. Dashed lines show results for the OAT echo with optimized readout direction such that the SNR is maximized. Dashed-dotted lines depict the case of an optimal alignment of the squeezed state before the phase imprint. For comparison, the results of the clean OAT echo and the TACT echo are shown again (solid lines). Computations are carried out for $N=10^3$ particles and $\theta=0.001$.
(a,b) Dependence on the interaction strength, showing the range of $0$-$10\,$dB squeezing for both schemes. Here, the echo ratio is set to $r=1.5$.
(c,d) Dependence on the echo ratio $r$ with initial squeezing of $-10\,$dB.}
	\label{fig9}		
\end{figure}
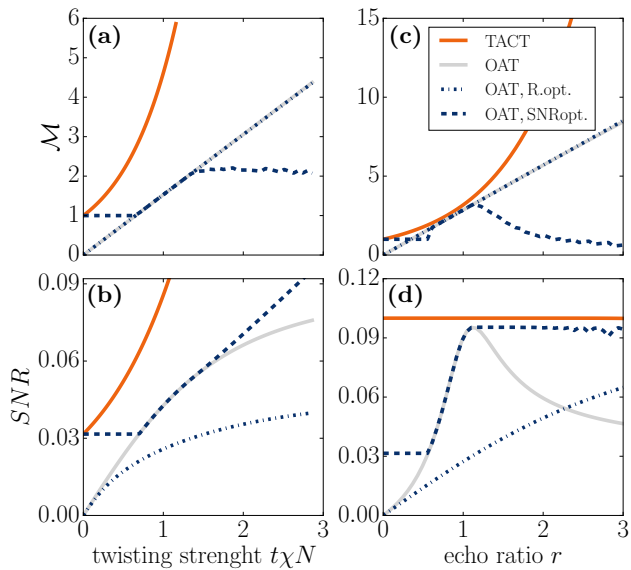

The OAT echo shows two main imperfections. First, it does not employ the initial entanglement completely since the squeezing ellipse is not optimally aligned. Second, the ideal readout direction is not fixed but dependent on the echo strength.
We extend the OAT echo scheme by two additional rotations of the quantum state, one before the phase imprint to align the squeezing ellipse and another one before the final detection to optimize the readout. 
The readout direction can either be optimized with respect to the SNR (proportional to the ideal metrological gain) or with respect to the magnification factor (proportional to the noise robustness). Thus optimizing the readout direction always leads to a trade-off between both.
Our analysis shows that both optimizations (alone or in combination) can lead to improved results for specific parameter regimes. However, they do not improve the performance of the OAT scheme in general, and never surpass the results of the TACT echo scheme.

\begin{figure}[t!]
	\centering
	\begin{tikzpicture}
	\node[inner sep=0pt] (left) at (0,0)
	    {\includegraphics[width=\columnwidth]{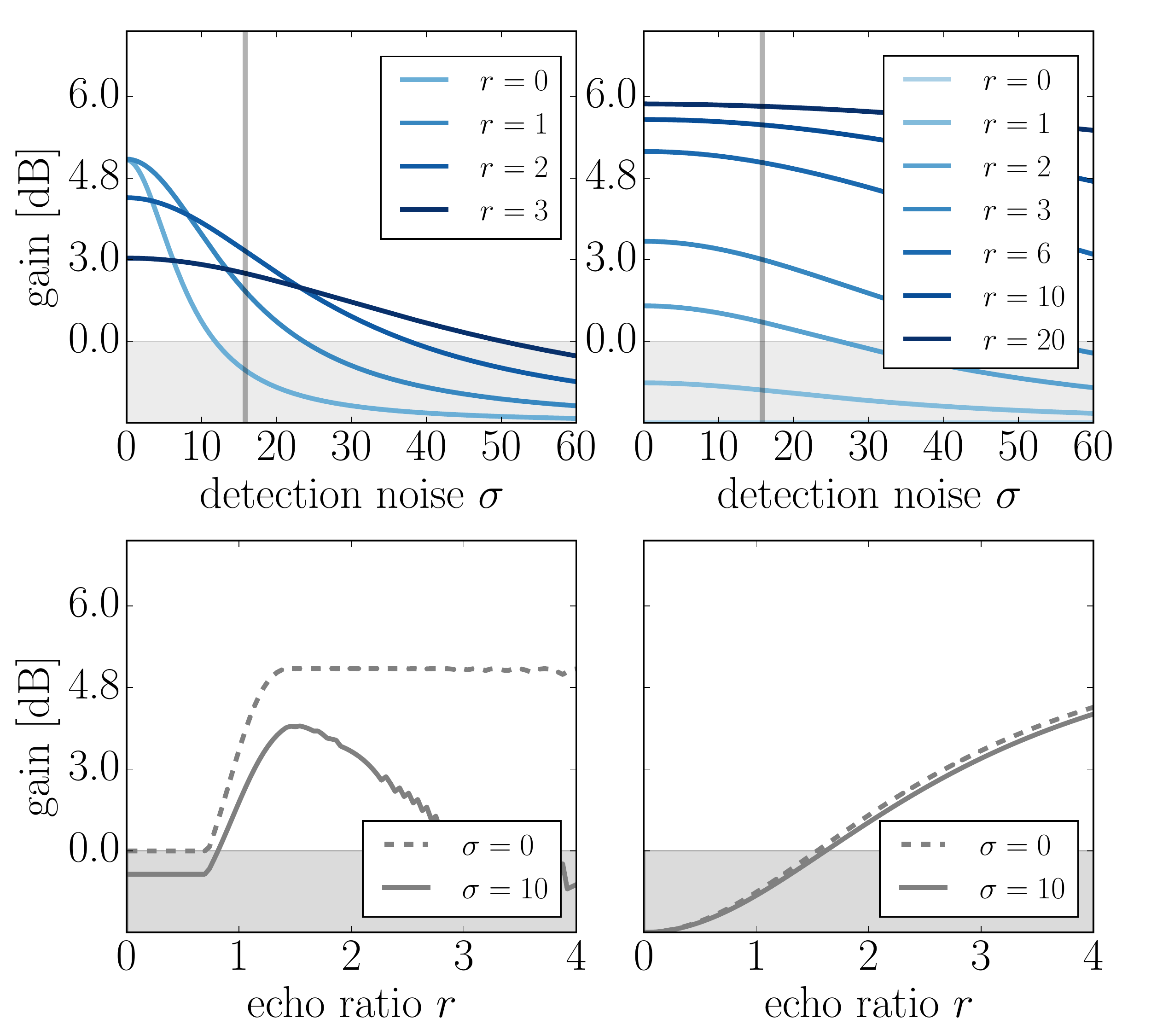}};
	\node[inner sep=0pt] (4) at (-3.05, 3.41) {\small \textbf{(a)}};
	\node[inner sep=0pt] (2) at (-3.05,-.43) {\small \textbf{(b)}};
	\node[inner sep=0pt] (1) at (.8,3.41) {\small \textbf{(c)}};
	\node[inner sep=0pt] (3) at (.8,-0.43) {\small \textbf{(d)}};
	\end{tikzpicture}
	\caption[]{(Color online) Metrological gain of the 
OAT echo with optimized readout direction (left panels) and the OAT echo with initial alignment of the squeezing (right panels), both with initial squeezing of $-6\, \mathrm{dB}$ and a particle number of $N=10^3$. 
Solid colored lines show numerical results for several echo ratios $r$. 
(a,c) Noise dependence of the gain. The gray vertical lines indicate the CSS noise level. The phase is set to an optimal small value.
(b,d) Dependence on the echo ratio for both optimization scenarios with vanishing detection noise (dashed line) and an exemplary noise of $\sigma=10$ (solid line).
}\label{fig10}		
\end{figure}
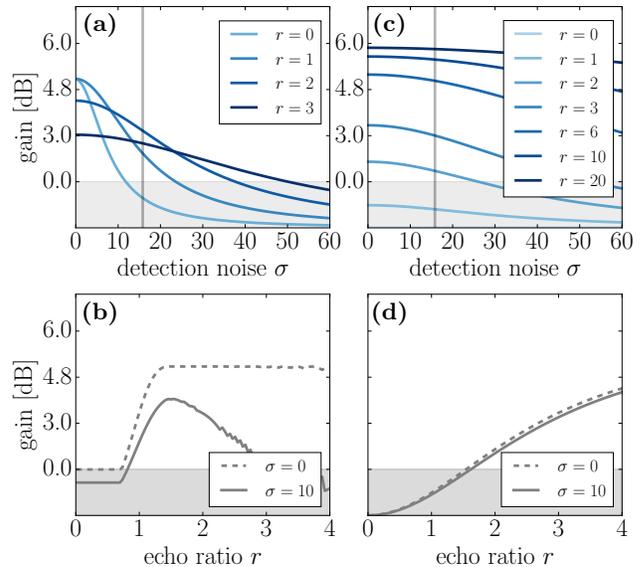

The following figures show the two most promising optimization scenarios, that is the initial alignment of the squeezing to the phase imprint and the optimization of the readout direction with respect to the SNR. Figure~\ref{fig9} visualizes both scenarios in terms of the SNR and the magnification factor. For comparison, we also plot the results of the clean OAT scheme and the TACT echo again.
The optimization of the readout direction with respect to the SNR (dashed black lines) improves the OAT echo for small twisting strengths and echo ratios.
For larger squeezing or stronger echo interactions, the SNR surpasses the original OAT scheme at the expense of a reduced magnification. This corresponds to a higher ideal metrological gain, but a smaller robustness to detection noise (Fig.~\ref{fig10}~(b))
If the alignment of the initial squeezing ellipse is optimized, the resulting magnification remains unchanged, while the SNR shows a much different behavior. Here, the performance is reduced in the regime of echo ratios up to $r \approx 2$. For much stronger echo interactions, however, the clean OAT echo can be overcome, finally approaching the performance of the TACT echo in the case of very large magnifications. As illustrated in Fig.~\ref{fig10}~(c), this optimized OAT echo can as well optimally exploit a given initial squeezing and fully preserve the associated metrological gain in the presence of strong detection noise (in agreement with the results of Ref. \cite{Hosten2016a}).

\section{Measurement in the effective basis}

Here we show details on the implementation of the effective two-mode measurement $(
N_0)$ and $(N_S + N_A)$. We perform numerical computations in the full three-mode system and ``crop'' the final distribution of outcomes according to the fact, that $N_S$ and $N_A$ cannot be determined separately but only measured as a sum. Therefore, all outcomes possessing the same number sum of particles in mode $(S)$ and $(A)$ are merged to one indistinguishable outcome, leading to a partial loss of information.

As our effective observable we define  $J_{z3} \defeq \frac{1}{2} [\hat{N}_0 - (\hat{N}_S + \hat{N}_A)]$, which becomes the common two-mode $J_z$ for $N_A \approx 0$.
Introducing the computational basis
\begin{align}
\ket{i;k} = \ket{N - (i+k) ,i ,k},
\end{align}
with $ \quad i,k = 0, 1, ..., N$ and $i+k \leq N$,
the spectral decompositions of the effective observables reads
\begin{align}
J_{z3} &= \sum_{i,k} \frac{N- 2(i+k)}{2} \ket{i;k}\bra{i;k}.
\end{align}
The incapability of separately determining the particle number in modes $(S)$ and $(A)$ manifests in degenerate eigenvalues for $J_{z3}$.
The full distribution $P_{ik} = |\braket{i;k|\psi}|^2$ cannot be obtained by this observable.
We obtain an appropriate, ``cropped" distribution by
$P_l = \sum_{\substack{i, k = 0 \\ i + k = l}}^l P_{ik} = \sum_{k=0}^l P_{l-k,k} $.
In this way, it only contains measurable information according to $J_{z3}$. The Fisher information based on this distribution describes the actual interferometric sensitivity more realistically.

\bibliography{main}
\end{document}